\documentclass[11pt,twoside,superscriptaddress,showkeys, showpacs]{revtex4}
\usepackage{graphicx, latexsym, amssymb, amsmath, color, multirow, mathrsfs, booktabs}

\usepackage[section]{placeins}

\newcommand{\beq}{\begin{equation}}
\newcommand{\eeq}{\end{equation}}
\newcommand{\beqn}{\begin{eqnarray}}
\newcommand{\eeqn}{\end{eqnarray}}
\newcommand{\bea}{\begin{array}}
\newcommand{\eea}{\end{array}}
\newcommand{\bsub}{\begin{subequations}}
\newcommand{\esub}{\end{subequations}}
\newcommand{\bpm}{\begin{pmatrix}}
\newcommand{\epm}{\end{pmatrix}}

\newcommand{\scr}[1]{{\mathscr #1}}
\newcommand{\ff}[1]{\frac{1}{#1}}
\newcommand{\lc}{\left<}
\newcommand{\rc}{\right>}
\newcommand{\lr}{\left|}
\newcommand{\rl}{\right|}

\newcommand{\lrb}[1]{\left(#1\right)}
\newcommand{\lrs}[1]{\left[#1\right]}

\newcommand{\Lrb}[1]{\left\{#1\right\}}

\newcommand{\svec}[1]{{\mbox{\boldmath${ #1}$}}}
\newcommand{\ivec}{\vec}

\newcommand{\secref}[1]{{\sf\bfseries Section \ref{#1}}}
\newcommand{\tabref}[1]{{\sf\bfseries Table \ref{#1}}}
\newcommand{\figref}[1]{{\sf\bfseries Fig. \ref{#1}}}

\newcommand{\nr}{{\text{NR}}}
\newcommand{\er}{{\text{R}}}

\begin{document} 

\title{Shell Structure and $\rho$-Tensor Correlations in Density-Dependent Relativistic Hartree-Fock theory}
\author{WenHui Long}\email{whlong@pku.org.cn}
 \affiliation{School of Physics, Peking University, 100871 Beijing, China}
 \affiliation{Center for Mathematical Sciences, University of Aizu, Aizu-Wakamatsu, 965-8580
Fukushima, Japan}
\author{Hiroyuki Sagawa}
 \affiliation{Center for Mathematical Sciences, University of Aizu, Aizu-Wakamatsu, 965-8580
Fukushima, Japan}
\author{Nguyen Van Giai}
 \affiliation{Institut de Physique Nucl\'eaire, CNRS-IN2P3, 
 F-91406 Orsay Cedex, France}
 \affiliation{Universit\'e Paris-Sud, F-91405 Orsay, France}
\author{Jie Meng}
 \affiliation{School of Physics, Peking University, 100871 Beijing, China}
 \begin{abstract}

A new effective interaction PKA1 with $\rho$-tensor couplings for the density-dependent relativistic Hartree-Fock
(DDRHF) theory is presented. It is obtained by fitting selected empirical ground state and shell structure properties.
It provides satisfactory descriptions of nuclear matter and the ground state properties of finite nuclei at the same
quantitative level as recent DDRHF and RMF models. Significant improvement on the single-particle spectra is also found
due to the inclusion of $\rho$-tensor couplings. As a result, PKA1 cures a common disease of the existing DDRHF and RMF
Lagrangians, namely the artificial shells at 58 and 92, and recovers the realistic sub-shell closure at $64$. Moreover,
the proper spin-orbit splittings and well-conserved pseudo-spin symmetry are obtained with the new effective
interaction PKA1.  Due to the extra binding introduced by the $\rho$-tensor correlations, the balance between the
nuclear attractions and the repulsions is changed and this constitutes the physical reason for the improvement of the
nuclear shell structure.

 \end{abstract}
\maketitle

\section{Introduction}

Within the relativistic scheme, mean field theory has achieved great success in the description of finite nuclei and
nuclear matter during the past years. One of the most outstanding models is the relativistic Hartree approach with the
no-sea approximation, namely the relativistic mean field (RMF) theory \cite{Miller:1972, Walecka:1974,Serot:1986}. The
RMF theory provides appropriate quantitative descriptions for both stable and exotic nuclei with a limited number of
free parameters, i.e., meson masses and meson-nucleon coupling constants \cite{Reinhard:1989, Ring:1996,
Lalazissis:1997, Serot:1997, Typel:1999,Niksic:2002, Bender:2003,Long04, Meng:2006}. Especially, the RMF model provides
a natural mechanism for explaining the spin-orbit splittings in nuclear spectra with the covariant formulation of the
strong scalar and vector fields. This feature becomes even more of central importance with the experimental observation
that nuclei near drip lines undergo a modification of their shell structure, where the spin-orbit potential must play
an essential role.

In the framework of the RMF approach, however, there exist two serious defects. One is the missing of one-pion exchange
process. Namely, with the relativistic Hartree approach, the one-pion exchange has zero contribution in mediating
nuclear interactions. Because of its small rest mass, one cannot simulate the one-pion exchange contributions with
zero-range limit. It is expected that the one-pion exchange may have minor effects in the spin-saturated system,
whereas in the unsaturated systems it plays an essential role in determining the isospin dependence of the shell
evolutions \cite{Long:2007}. Another problem is the tensor correlations, e.g., the $\rho$-tensor couplings. In the
Hartree approximation, the contributions from the tensor couplings are practically negligible. In recent
non-relativistic \cite{Otsuka:2005, Colo:2007} and relativistic studies \cite{Long:2007}, it is shown that the tensor
forces have distinct effects on the shell evolution of nuclei \cite{Schiffer:2004}. In fact these two defects are
mainly due to the absence of Fock terms which are dropped in RMF.

During the past decades, there have been several attempts to include the Fock terms in the relativistic description of
nuclear systems \cite{Bouyssy:1985,Bouyssy:1987,Bernardos:1993,Marcos:2004}. These relativistic Hartree-Fock (RHF)
approaches could not provide satisfactory quantitative descriptions for the nuclear structure properties compared to
the RMF approach. This is mainly due to the numerical complexity induced by the inclusion of Fock terms, which strongly
increase the difficulties to find appropriate effective Lagrangians for the RHF approach. Recently a new RHF method,
the density-dependent relativistic Hartree-Fock (DDRHF) theory \cite{Long:2006} has brought a new insight to this
problem. With the effective Lagrangians of Refs. \cite{Long:2006, Long:2006a}, the DDRHF theory can describe the ground
state properties of many nuclear systems quantitatively on the same level as RMF. In addition, the investigations about
the nuclear shell structure evolution within the DDRHF theory indicate that the one-pion exchange has a significant
effect on the isospin dependence of the shell evolution \cite{Long:2007}.

However, some artificial shell structures, e.g., $Z=58$ and $Z=92$ appear in the calculations of RMF \cite{Geng:2006}
as well as RHF. These spurious shells lead to the overbinding problem in these regions \cite{Geng:2005} and they also
affect strongly the isospin dependence of the shell evolutions \cite{Long:2007}. The relative positions of $1g_{7/2}$
and $2d_{5/2}$ states induce an artificial shell closure at $Z =58$. The corresponding states for $Z=92$ are $1h_{9/2}$
and $2f_{7/2}$. As a common feature, all these states are high-$j$ states. Then, the single-particle energies of these
states will be strongly affected by the tensor force \cite{Long:2007}, e.g., $\rho$-tensor couplings, which was not
included in RMF or DDRHF before. In order to solve this artificial shell structure problem, we consider the
$\rho$-tensor correlations in this work. In \secref{Sec:DDRT}, we introduce the $\rho$-tensor couplings in the DDRHF
theory, where a new effective interaction PKA1 with the $\rho$-tensor coupling is presented. In \secref{sec:Structure},
the detailed investigation of the nuclear structure is performed with the newly obtained effective interaction PKA1.
The conclusions are drawn in \secref{sec:conc}.

\section{DDRHF theory with $\rho$-Tensor Coupling}\label{Sec:DDRT}
The starting point of the DDRHF theory is the Lagrangian associated with nucleon ($\psi$), isoscalar $\sigma$- and
$\omega$-mesons, isovector $\rho$- and $\pi$-mesons, and photon ($A$) fields \cite{Long:2006,Long:2006a}. In the
isoscalar channels, the $\sigma$-scalar and $\omega$-vector couplings provide the main part of the nuclear
interactions, i.e., the short-range repulsive and mid-, long-range attractive interactions, respectively. {One of the
distinct differences of RHF from RMF is that all the mesons, including the isoscalar ones, have significant
contributions to the isospin part of nuclear interactions.} In Refs. \cite{Long:2006,Long:2006a,Long:2006PS,Long:2007},
the $\rho$-vector and $\pi$-pseudo-vector couplings were introduced in the calculations of DDRHF. The recent
investigation about the role of one-pion exchange in DDRHF shows that the tensor type force has the strong effects on
the nuclear structure \cite{Long:2007}.  In this study, we introduce the $\rho$-tensor correlations into the DDRHF
theory in order to have a better understanding of the shell structure and cure the artificial shell structure problems
of DDRHF and RMF \cite{Geng:2006}.

\subsection{General Formalism for $\rho$-Nucleon Couplings}
In the DDRHF theory, the part of the Lagrangian containing the $\rho$-meson fields can be written as
 \beq\label{Lagrangian-NR}
\scr L_\rho =  - \ff4\ivec R_{\mu\nu}\cdot\ivec R^{\mu\nu} + \ff2 m_\rho^2 \ivec\rho_\mu\cdot\ivec\rho^\mu-g_\rho
\bar\psi\gamma_\mu\ivec\rho^\mu\cdot\ivec\tau\psi {+}
\frac{f_\rho}{2M}\bar\psi\sigma_{\mu\nu}\partial^\nu\ivec\rho^\mu\cdot\ivec\tau\psi~,
 \eeq
where $\ivec R_{\mu\nu} = \partial_\mu\ivec\rho_\nu -
\partial_\nu\ivec \rho_\mu$, and $m_\rho$ denotes the rest mass of
the isovector-vector $\rho$-meson ($\ivec\rho_\nu$), and $g_\rho$ and $f_\rho$ are the vector and tensor coupling
strengths, respectively.

From the Lagrangian (\ref{Lagrangian-NR}), one can obtain the equation of motion for the $\rho$-meson field as,
 \beq
\lrs{\square + m_\rho^2}\ivec\rho_\nu = g_\rho\bar\psi\gamma_\nu\ivec\tau\psi {+}
\partial^\mu\frac{f_\rho}{2M}\bar\psi\sigma_{\nu\mu}\ivec\tau\psi~,
 \eeq
which leads to the general form of $\rho$-meson field,
 \beq\label{RHO}
\ivec\rho_\nu(x) =  \int d^4y \lrs{ g_\rho\bar\psi\gamma_\nu\ivec\tau\psi {-}
\frac{f_\rho}{2M}\bar\psi\sigma_{\nu\mu}\ivec\tau\psi\partial^\mu }_{y} D_\rho(x, y)
 \eeq
where $D_\rho$ is the $\rho$-meson propagator.

With the Legendre transformation ($\phi_i$ is the field variables of the Lagrangian $\scr L$),
 \beq
\scr H_{\rho} = \frac{\partial\scr L_\rho}{\partial \dot\phi_i}\dot\phi_i - \scr L_\rho,
 \eeq
the Hamiltonian of the $\rho$-meson field can be obtained as,
 \beq
H_\rho =\ff2 \int d^3x \lrs{g_\rho\bar\psi\gamma^\nu\ivec\tau\psi {-} \frac{f_\rho}{2M}\bar\psi\sigma^{\nu
l}\ivec\tau\psi\partial_l}_x\centerdot\ivec\rho_\nu(x)~.
 \eeq
In the above expression, we neglect the time component of the four-momentum carried by the mesons, which amounts to
ignoring the retardation effects.

With the quantization of the nucleon field $\psi$
\cite{Bouyssy:1987} and the general form of the $\rho$-meson field
(\ref{RHO}), the Hamiltonian $H_\rho$ can be written as,
 \beq\begin{split}\label{Hrho}
H_\rho^{(i)} = \ff2\sum_{\alpha\beta\gamma\delta} c_\alpha^\dag c_\beta^\dag c_\gamma c_\delta& \lc\tau_\alpha\rl
\ivec\tau_1\lr\tau_\delta\rc\cdot\lc\tau_\beta\rl\ivec\tau_2\lr\tau_\gamma\rc\\&\times\int d\svec r_1 d\svec r_2\bar
f_\alpha(\svec r_1)\bar f_\beta(\svec r_2)\lrs{\Gamma_\rho^i(1,2)v(m_\rho;1, 2)} f_\gamma(\svec r_2)f_\delta(\svec
r_1)~,
 \end{split}\eeq
where the Yukawa factor $v(m_\rho;1,2)$ is
 \beq\label{Yukawa_rho}
v(m_\rho;1,2) = \frac{1}{4\pi}\frac{e^{-m_\rho\lr\svec r_1-\svec r_2\rl}}{\lr\svec r_1-\svec r_2\rl}~.
 \eeq
The inclusion of $\rho$-tensor correlations leads to three types of interactions in Eq. (\ref{Hrho}): the Vector (V),
Tensor (T) and Vector-Tensor (VT) couplings.  The corresponding vertex matrix $\Gamma_\rho^i(1,2)$ in Eq. (\ref{Hrho})
can be expressed as,
 \bsub\beqn
\Gamma_\rho^V(1,2) &=& g_\rho(1) \gamma_\mu g_\rho(2) \gamma^\mu(2),\\
\Gamma_\rho^T(1,2) &=& \frac{1}{4M^2}f_\rho(1)\sigma_{\nu k}(1)f_\rho(2)\sigma^{\nu
l}(2)\partial^k(1)\partial_l(2),\\
\Gamma_\rho^{VT}(1,2)&=& \ff{2M}\lrs{ f_\rho(1)\sigma^{{k\nu}}(1) g_\rho(2)\gamma_\nu(2)\partial_k(1) +
g_\rho(1)\gamma_\nu f_\rho(2)\sigma^{{k\nu}}(2)\partial_k(2)}.
 \eeqn\esub
Detailed expressions about the energy functional of $\rho$-meson field can be found in Refs. \cite{Bouyssy:1987,
Long:2006}.

In DDRHF, the meson-nucleon coupling constants are treated as a function of baryonic density $\rho_b$. Here we take the
same functional form for the density-dependence  of the isoscalar mesons ($g_\sigma$ and $g_\omega$) as in Ref.
\cite{Long:2006}:
 \beq
g_i(\rho_b) = g_i(\rho_0) f_i(\xi),~~~~~ \text{for } i = \sigma, \omega,
 \eeq
where
 \beq
f_i(\xi) = a_i \frac{1+b_i(\xi + d_i)^2}{1+c_i(\xi + d_i)^2}
 \eeq
with $\xi = \rho_b/\rho_0$, and $\rho_0$ denotes the saturation density of nuclear matter. For the isovector mesons as
well as the newly introduced $\rho$-tensor coupling $f_\rho$, the exponential density-dependence is adopted as,
 \beq
g_i = g_i(0) e^{-a_i \ \xi}~.
 \eeq
In the above expression, $g_i(0)$ corresponds to the free coupling constants $g_\rho, f_\rho$, and $f_\pi$, and $a_i$
are the corresponding parameters $a_\rho$, $a_T$ and $a_\pi$, respectively.

For the open shell nuclei, the pairing correlations are treated by the BCS method and the pairing matrix elements are
calculated with a zero-range, density-dependent interaction \cite{Dobaczewski:1996}
 \beq
V(r_1, r_2) = V_0 \delta(r_1-r_2) \lrs{ 1- \frac{\rho_b(r)}{\rho_0}}~,
 \eeq
where $V_0$ = -900 MeV$\cdot$fm$^3$. The active pairing space is limited to the single-particle states below the
single-particle energy +15 MeV.

In this work, the corrections from the center-of-mass motion are treated in the same way as in Refs. \cite{Long04,
Long:2006a}. For the numerical calculations, a box boundary condition at $20$fm is introduced for the unbound states as
well as the bound ones and we check that the overall results are not affected by the choice of the box size. For the
radial step, one may choose smaller one about 0.05 fm in the light nuclei whereas 0.1 fm is precise enough for the
heavy nuclei.

\subsection{New Effective Interaction}
In the previous DDRHF parametrizations (PKO1, PKO2 and PKO3) \cite{Long:2006, Long:2006a, Long:2007}, we selected 12
nuclei as the reference ones, i.e., $^{16}$O, $^{40}$Ca, $^{48}$Ca, $^{56}$Ni, $^{68}$Ni, $^{90}$Zr, $^{116}$Sn,
$^{132}$Sn, $^{182}$Pb, $^{194}$Pb, $^{208}$Pb and $^{214}$Pb. In the present case for PKA1, we aim to cure the
artificial shell structure problem at $Z = 58$ and $92$. To solve this problem and recover the sub-shell closure at $Z
= 64$, two more nuclei $^{140}_{~58}$Ce$_{82}$ and $^{146}_{~64}$Gd$_{82}$ are added as the reference nuclei and
$^{56}$Ni is replaced by its neighboring one $^{58}$Ni. The parameter fitting procedure is similar to that in Refs.
\cite{Long04,Long:2006a}. Besides the bulk properties ($\rho_0$, $K$ and $J$) of nuclear matter and the binding
energies of the reference nuclei, we include the spin-orbit splittings of neutron and proton $1p$ states of $^{16}$O,
and the shell gaps at $Z=58$ ($^{140}_{~58}$Ce$_{82}$) and $Z=64$ ($^{146}_{~64}$Gd$_{82}$) as the new criteria. By
minimizing the $\chi^2$ error as in Ref. \cite{Long:2006a}, we obtain the new effective interaction PKA1 with the
$\rho$-tensor coupling for the DDRHF theory (see \tabref{tab:PKA1}). In this parametrization, we have 12 free
parameters, 6 in the isoscalar channels as well as 6 in the isovector channels. We slightly change the coupling
strength for $\pi$-meson ($f_\pi$ and $a_\pi$) in PKA1 from the effective interaction PKO1 \cite{Long:2006}, which is
the starting Lagrangian in the fitting process.

\begin{table}[htbp]\setlength{\tabcolsep}{6pt}
\caption{Parameters of new effective interaction PKA1. The quantity $\kappa$ is $\kappa = f_\rho/g_\rho$, the ratio of
the vector and tensor couplings of $\rho$-meson. The masses (in MeV) of nucleon, $\omega$-, $\rho$- and $\pi$-mesons
are taken as $M = 938.9$, $m_\omega = 783.0, m_\rho = 769.0$ and $m_\pi = 138.0$, respectively.}\label{tab:PKA1}
 \begin{tabular}{c|r|c|r|c|r}\toprule[1.5pt]\toprule[0.5pt]
 $m_\sigma$&488.227904&$a_\sigma$& 1.103589& $a_\omega$&  1.126166\\
 $g_\sigma$&  8.372672&$b_\sigma$&16.490109& $b_\omega$&  0.108010\\
 $g_\omega$& 11.270457&$c_\sigma$&18.278714& $c_\omega$&  0.141251\\
 $\rho_0$  &  0.159996&$d_\sigma$& 0.135041& $d_\omega$&  1.536183\\ \midrule[0.5pt]
 $g_\rho$  &  3.649857&$\kappa$  & 3.199491& $f_\pi$   &  1.030722\\
 $a_\rho$  &  0.544017&$a_T$     & 0.820583& $a_\pi$   & 1.200000\\
\bottomrule[0.5pt]\bottomrule[1.5pt]
\end{tabular}
\end{table}

\begin{figure}[htbp]
\includegraphics[width = 0.4\textwidth]{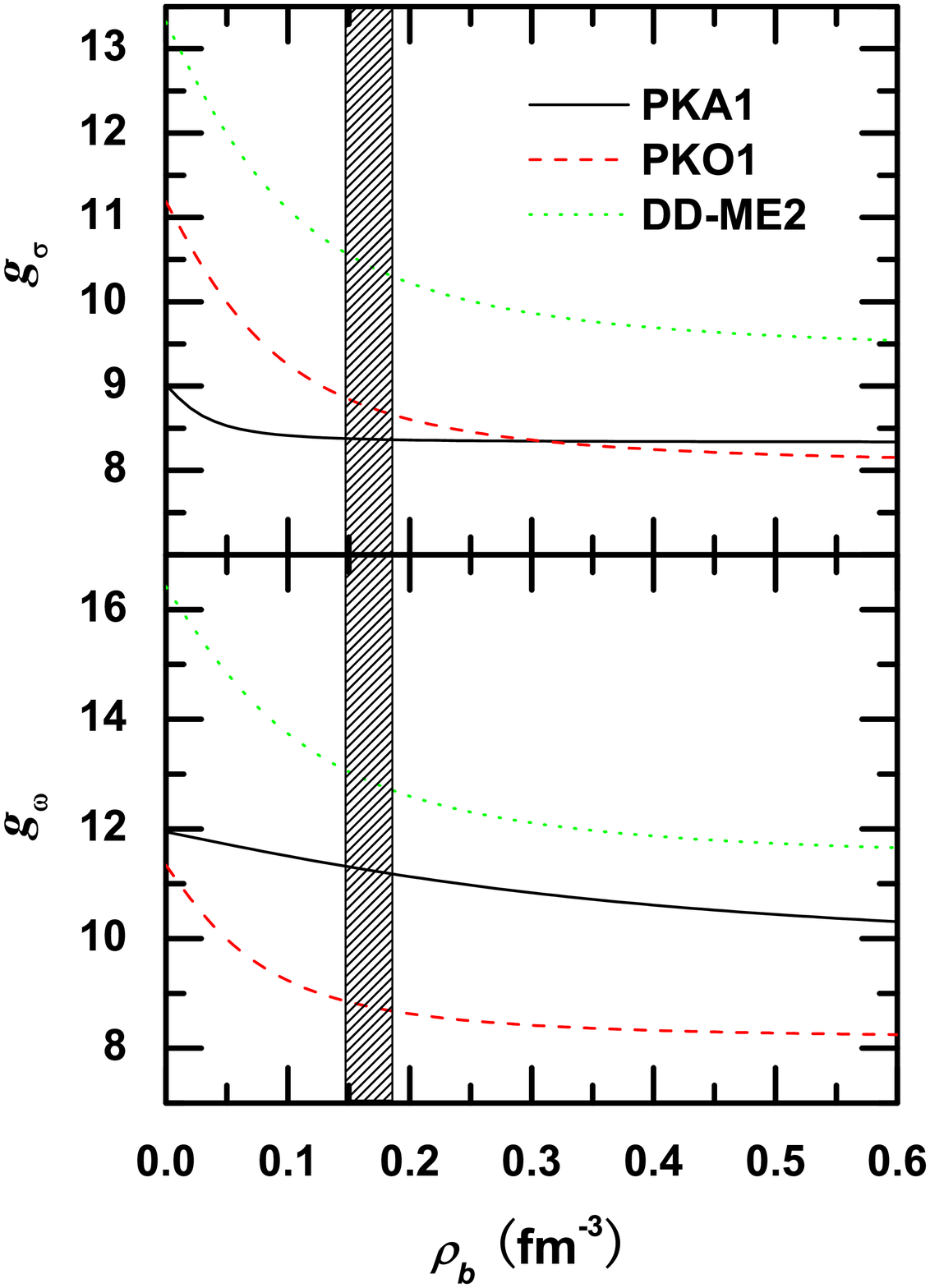}
\includegraphics[width = 0.4\textwidth]{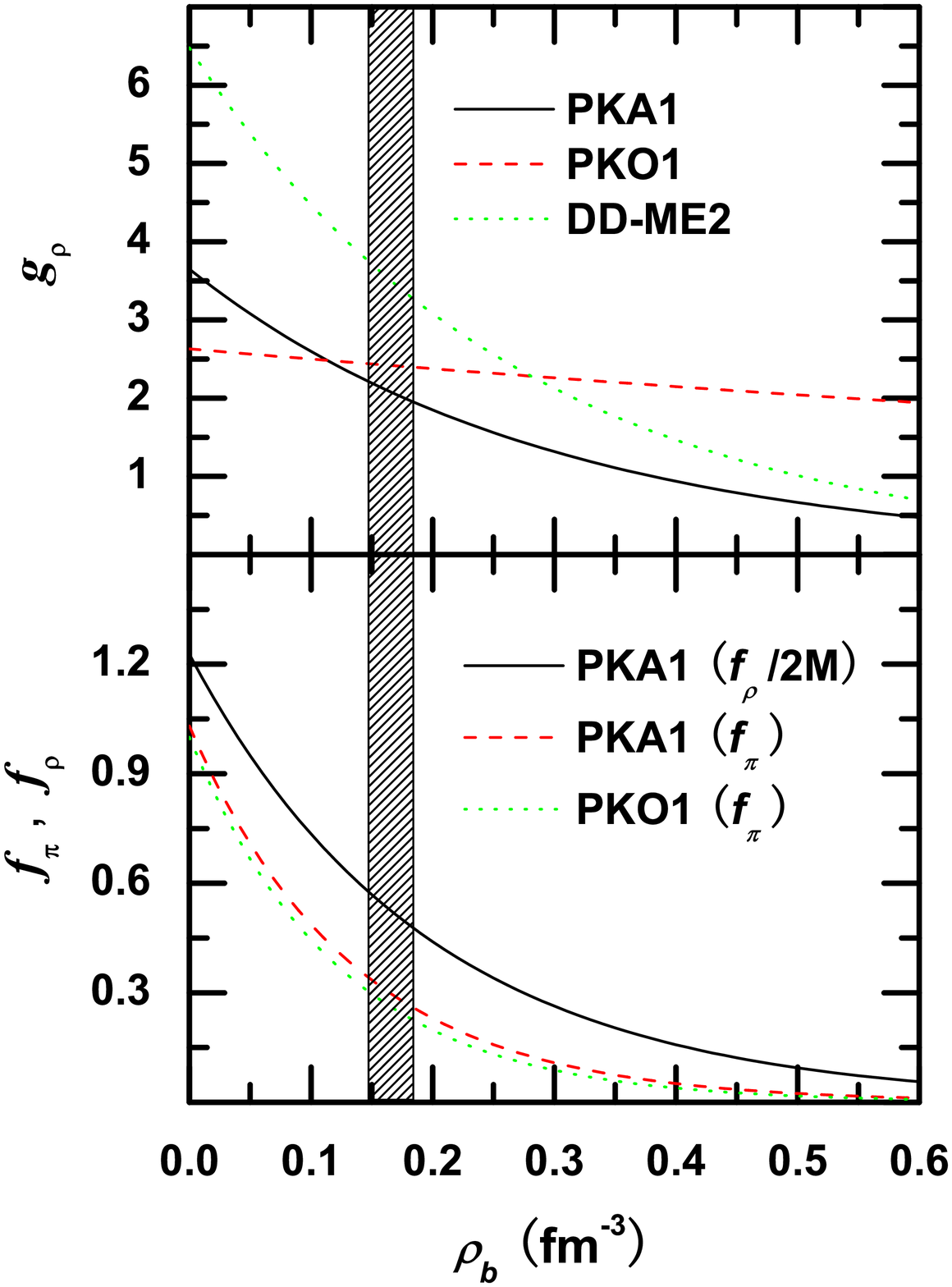}
\caption{(color online) The density-dependent meson-nucleon couplings in the isoscalar (left panel: $g_\sigma$ and
$g_\omega$) and isovector (right panel: $g_\rho$, $f_\rho$ and $f_\pi$) channels as functions of density for the new
DDRHF effective interaction PKA1, in comparison with PKO1 in DDRHF and DD-ME2 in RMF. The shadowed area denotes the
empirical saturation density region.}\label{fig:couplings}
 \end{figure}

In \figref{fig:couplings}, the density-dependent couplings $g_\sigma$, $g_\omega$ (left panels) and $g_\rho$, $f_\rho$,
and $f_\pi$ (right panels) of the effective interaction PKA1 are shown as functions of the baryonic density $\rho_b$,
in comparison with PKO1 (DDRHF) \cite{Long:2006} and DD-ME2 (RMF) \cite{Lalazissis:2005}. As seen from
\figref{fig:couplings}, the density-dependence of isoscalar couplings, especially for $g_\sigma$, is weak for PKA1
compared to PKO1 and DD-ME2. For the isovector channels, $g_\rho$, $f_\rho$ and $f_\pi$ show strong density-dependence
in PKA1 whereas the density-dependence of $g_\rho$ of PKO1 is weak. For the one-pion exchange, PKA1 and PKO1 have
nearly the same coupling strength for $f_\pi$. PKA1 and PKO1 have smaller coupling strength in magnitude for both the
isoscalar and isovector channels compared to DD-ME2. This is mainly due to the effects of the Fock terms. One can also
find that the inclusion of $\rho$-tensor coupling leads to smaller $g_\sigma$ and larger $g_\omega$ in PKA1 than those
in PKO1. This indicates that the $\rho$-tensor correlations contribute to make the nuclear interactions attractive.

We calculate the bulk properties of nuclear matter with the effective interaction PKA1 as shown in \tabref{tab:nucm},
where the results calculated by DDRHF with PKO1 and RMF with DD-ME2 are also listed for comparison. Compared to PKO1
and DD-ME2, PKA1 gives a larger saturation density $\rho_0$, which is close to the common value in non-relativistic HF
calculations. Although PKA1 gives a smaller compression modulus $K$ and a larger symmetry energy $J$, the values are
still acceptable.  Among these three effective interactions, PKO1 has the largest effective masses (the relativistic
one $M_\er^*$ and the non-relativistic one $M_\nr^*$) \cite{Long:2006} whereas DD-ME2 gives smallest ones. Comparing
the values of $M_\nr^*$ and $M_S^*$, one can find a significant difference between DDRHF and RMF, the DDRHF models
giving a larger difference between these two masses. Actually this difference is of special importance in describing
nuclear structure since the effective mass $M_\nr^*$ is related to the level density whereas the scalar mass $M_S^*$ is
related to the spin-orbit splitting. In the following, one may find the corresponding effects due to this difference.

\begin{table}[htbp]\setlength{\tabcolsep}{6pt}
\caption{The saturation density $\rho_0$ (fm$^{-3}$), the binding energy $E_B$ (MeV), the compression modulus $K$
(MeV), the symmetry energy $J$ (MeV), the effective masses $M_\er^*$ and $M_\nr^*$ \cite{Long:2006}, and the scalar
mass $M_S^*$ for the symmetric nuclear matter. The results are calculated with PKA1 and PKO1 in DDRHF, and with DD-ME2
in RMF.}\label{tab:nucm}
 \begin{tabular}{c|cccc|ccc}\toprule[1.5pt]\toprule[0.5pt]
       &$\rho_0$& $E_B$  & $K$    &  $J$  &$M_\er^*$&$M_\nr^*$&$M_S^*$\\ \midrule[0.5pt]
PKA1   & 0.160  & -15.83 & 229.96 & 36.02 & 0.663   & 0.681   & 0.547 \\
PKO1   & 0.152  & -16.00 & 250.24 & 34.37 & 0.727   & 0.746   & 0.590 \\
DD-ME2 & 0.152  & -16.14 & 250.97 & 32.30 & 0.635   & 0.652   & 0.572 \\
\bottomrule[0.5pt]\bottomrule[1.5pt]
\end{tabular}
\end{table}

As we have mentioned before, 14 reference nuclei are adopted in the parametrization of PKA1. The binding energies and
charge radii of these nuclei calculated with PKA1 are shown in \tabref{tab:observ} in comparison with the calculations
of PKO1 \cite{Long:2006}, DD-ME2 \cite{Lalazissis:2005}, and the experimental data \cite{Audi:2003, Vries:1987,
Nadjakov:1994}. Other reference nuclei than the present 14 ones used in the parametrizations PKO1 and DD-ME2 are listed
in the lower panel of \tabref{tab:observ}. In this table, the root mean square deviations (rmsd) $\Delta$ and relative
rmsd $\delta$ are defined as
 \bsub\begin{align}
\Delta^2 \equiv &{\ff N\sum_{i=1}^N {\lrb{y_i^{\text{Exp.}} - y_i^{\text{Cal.}}}^2}}~,\\
\delta^2 \equiv &{\ff N\sum_{i=1}^N {\lrb{1 - y_i^{\text{Cal.}}/y_i^{\text{Exp.}}}^2}}~.
 \end{align}\esub
From the values of $\Delta$ and $\delta$ in \tabref{tab:observ}, we can see that PKA1 provides an appropriate
description for both binding energies and charge radii of these reference nuclei. For the binding energies, PKA1 gives
the best agreement with the data. For the charge radii, PKA1 gives quantitatively comparable description to PKO1 and
DD-ME2. {Since the reference nuclei cover from light ($^{ 16}$O) to heavy ($^{214}$Pb, $^{210}$Po) ones, one may expect
that DDRHF with the tensor interaction PKA1 can provide a proper description of the nuclei in the whole nuclear chart.}

\begin{table}[htbp]\setlength{\tabcolsep}{6pt}
\caption{Binding energies and charge radii for the reference nuclei. The results are calculated by DDRHF with the new
effective interaction PKA1 and with PKO1 \cite{Long:2006}, and by RMF with DD-ME2 \cite{Lalazissis:2005}. The upper
panel gives the reference nuclei of PKA1 and the other reference nuclei of PKO1 and DD-ME2 are listed in the lower
panel. Experimental data are taken from Refs. \cite{Audi:2003, Vries:1987, Nadjakov:1994}.}\label{tab:observ}
 \begin{tabular}{c|rrrr|rrrr}\toprule[1.5pt]\toprule[0.5pt]
            &\multicolumn{4}{|c|}{$E_b$(MeV)}                    &\multicolumn{4}{c}{$r_{\text{ch}}$(fm)}\\
 Nuclide    &     Exp.~~  &    PKA1~~  &    PKO1~~  &  DD-ME2    &  Exp.   & PKA1    &  PKO1  & DD-ME2   \\ \midrule[0.5pt]
 $^{ 16}$O  &  -127.6193  & -126.9913  & -128.3250  & -127.9640  &  2.7370 &  2.7996 & 2.6757 & 2.6703   \\
 $^{ 40}$Ca &  -342.0520  & -341.7177  & -343.2747  & -343.0046  &  3.4852 &  3.5251 & 3.4423 & 3.4417   \\
 $^{ 48}$Ca &  -415.9904  & -416.3697  & -417.3713  & -414.9174  &  3.4837 &  3.4916 & 3.4501 & 3.4568   \\
 $^{ 58}$Ni &  -506.4584  & -505.9705  & -502.9424  & -501.0959  &  3.7827 &  3.7006 & 3.7195 & 3.7387   \\
 $^{ 68}$Ni &  -590.4077  & -590.1520  & -591.4448  & -591.6165  &         &  3.8766 & 3.8480 & 3.8651   \\
 $^{ 90}$Zr &  -783.8919  & -784.3525  & -784.6039  & -782.4711  &  4.2720 &  4.2794 & 4.2501 & 4.2574   \\
 $^{116}$Sn &  -988.6835  & -986.9107  & -987.7642  & -986.8494  &  4.6257 &  4.6056 & 4.5924 & 4.6061   \\
 $^{132}$Sn &  -1102.8508 & -1103.2468 & -1103.5838 & -1102.9335 &         &  4.6985 & 4.6964 & 4.7047   \\
 $^{140}$Ce &  -1172.6915 & -1170.0999 & -1177.5650 & -1175.3038 &  4.8774 &  4.8836 & 4.8672 & 4.8690   \\
 $^{146}$Gd &  -1204.4353 & -1202.0964 & -1205.0765 & -1203.2093 &  4.9838 &  4.9889 & 4.9669 & 4.9771   \\
 $^{182}$Pb &  -1411.6534 & -1409.7315 & -1412.7060 & -1411.0088 &         &  5.3831 & 5.3708 & 5.3819   \\
 $^{194}$Pb &  -1525.8907 & -1521.9706 & -1523.8836 & -1522.1179 &  5.4446 &  5.4488 & 5.4334 & 5.4431   \\
 $^{208}$Pb &  -1636.4301 & -1636.9604 & -1636.9108 & -1638.0676 &  5.5046 &  5.5103 & 5.5051 & 5.5092   \\
 $^{214}$Pb &  -1663.2906 & -1661.3564 & -1662.4803 & -1659.5703 &  5.5622 &  5.5600 & 5.5619 & 5.5605   \\ \midrule[0.5pt]
 $\Delta$   &             &     1.6847 &     1.8787 &     2.3495 &         &  0.0342 & 0.0341 & 0.0298   \\
 $\delta$   &             &     0.19\% &     0.30\% &     0.34\% &         &  1.03\% & 1.01\% & 0.94\%   \\ \midrule[0.5pt]\midrule[0.5pt]
 $^{ 56}$Ni &  -483.9917  & -486.2191  & -483.0607  & -481.1788  &         &  3.6662 & 3.6899 & 3.7114   \\
 $^{ 72}$Ni &  -613.1694  & -613.8442  & -615.1913  & -613.1482  &         &  3.9029 & 3.8795 & 3.8964   \\
 $^{124}$Sn &  -1049.9627 & -1049.9471 & -1050.1779 & -1049.0708 &  4.6739 &  4.6649 & 4.6477 & 4.6602   \\
 $^{204}$Pb &  -1607.5059 & -1605.8114 & -1606.7831 & -1606.4818 &  5.4861 &  5.4936 & 5.4857 & 5.4908   \\
 $^{210}$Po &  -1645.2125 & -1644.7950 & -1646.4921 & -1646.9803 &         &  5.5482 & 5.5393 & 5.5425   \\ \midrule[0.5pt]
 $\Delta$   &             &     1.5927 &     1.7256 &     2.1784 &         &  0.0317 & 0.0322 & 0.0277   \\
 $\delta$   &             &     0.20\% &     0.27\% &     0.32\% &         &  0.95\% & 0.94\% & 0.87\%   \\
\bottomrule[0.5pt]\bottomrule[1.5pt]
\end{tabular}
\end{table}

\begin{figure}[htbp]
\includegraphics[width = 0.6\textwidth]{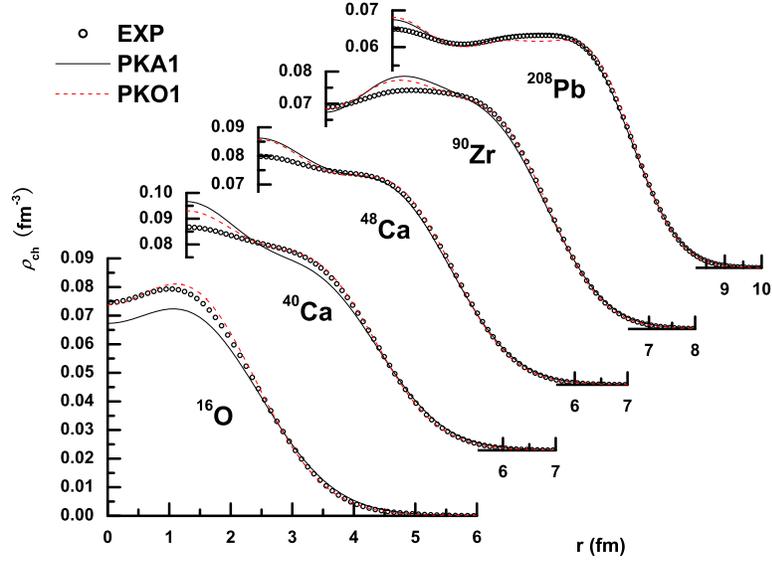}
\caption{(color online) Charge distributions of $^{16}$O, $^{40}$Ca, $^{48}$Ca, $^{90}$Zr and $^{208}$Pb. The results
are calculated with PKA1 and PKO1. Experimental data are taken from Ref. \cite{Vries:1987}. }\label{fig:Density}
 \end{figure}

The charge densities calculated with PKA1 and PKO1 for the nuclei $^{16}$O, $^{40}$Ca, $^{48}$Ca, $^{90}$Zr and
$^{208}$Pb are presented in \figref{fig:Density}. The experimental data \cite{Vries:1987} are also shown for
comparison. One can see that DDRHF with PKA1 provides a fairly good agreement with the data in heavy systems, e.g.,
$^{208}$Pb and $^{90}$Zr. For $^{40}$Ca and $^{48}$Ca PKA1 also shows comparable quality to PKO1 whereas it presents
less good agreement for $^{16}$O. In \tabref{tab:observ}, one can also find better agreement in the heavy nuclei than
in the light ones for the charge radii calculated with PKA1.

\section{Nuclear Structure Properties with Tensor Correlations}\label{sec:Structure}

\subsection{Single-particle Spectra}
In \secref{Sec:DDRT}, we showed that the effective interaction PKA1 can describe quantitatively well the bulk
properties of nuclear matter and the ground state properties of finite nuclei. In this section, we study the shell
structure of the Hartree-Fock single-particle energies for several reference nuclei by using PKA1. {For comparison, we
also show the results of the reference nuclei calculated with three other interactions: PKO1 (DDRHF without
$\rho$-tensor couplings), PK1 \cite{Long04} (RMF with non-linear self-couplings of mesons), DD-ME2 (RMF with
density-dependent meson-nucleon couplings).}

\begin{figure}[htbp]
\includegraphics[width = 0.40\textwidth]{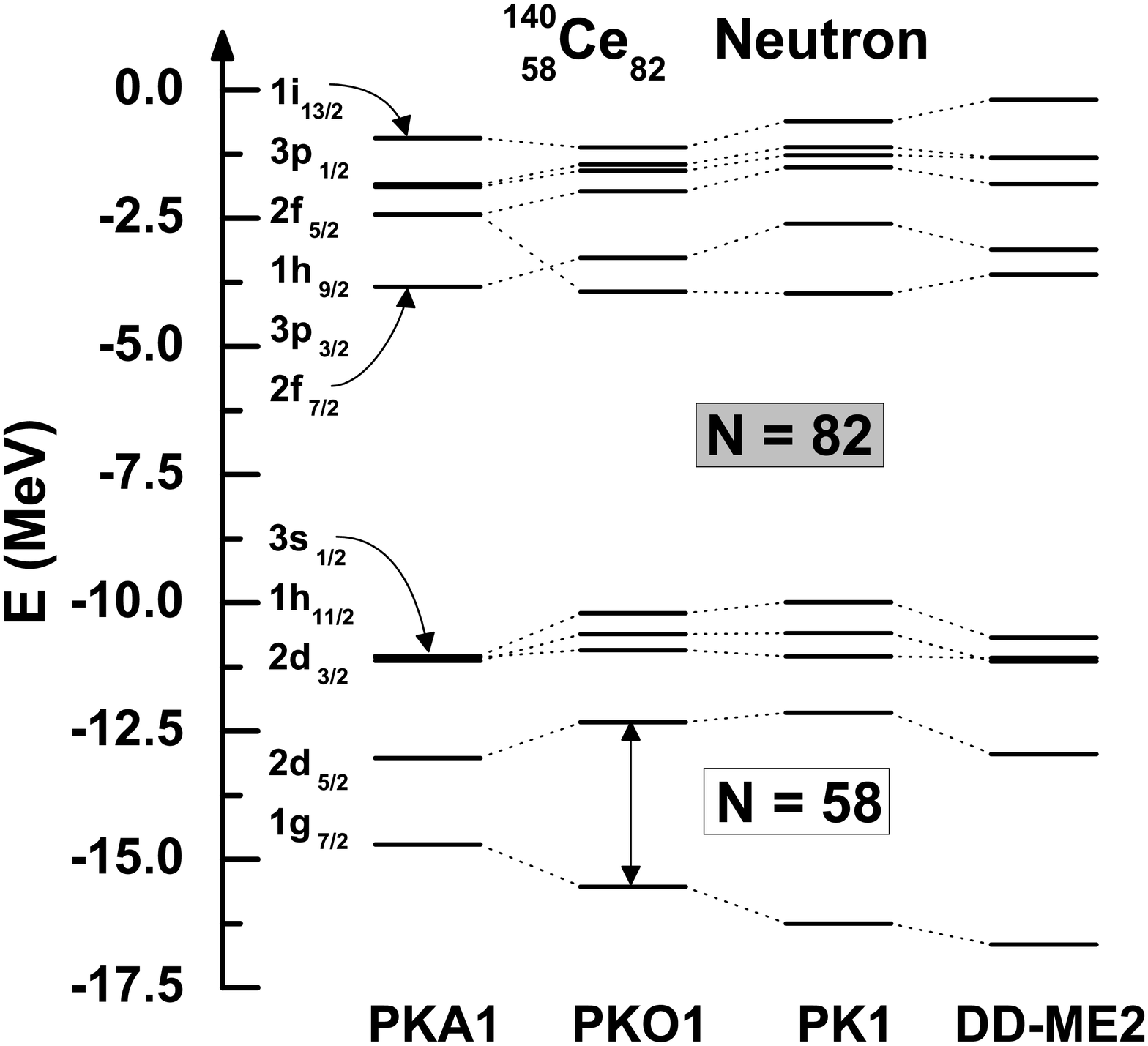}
\includegraphics[width = 0.40\textwidth]{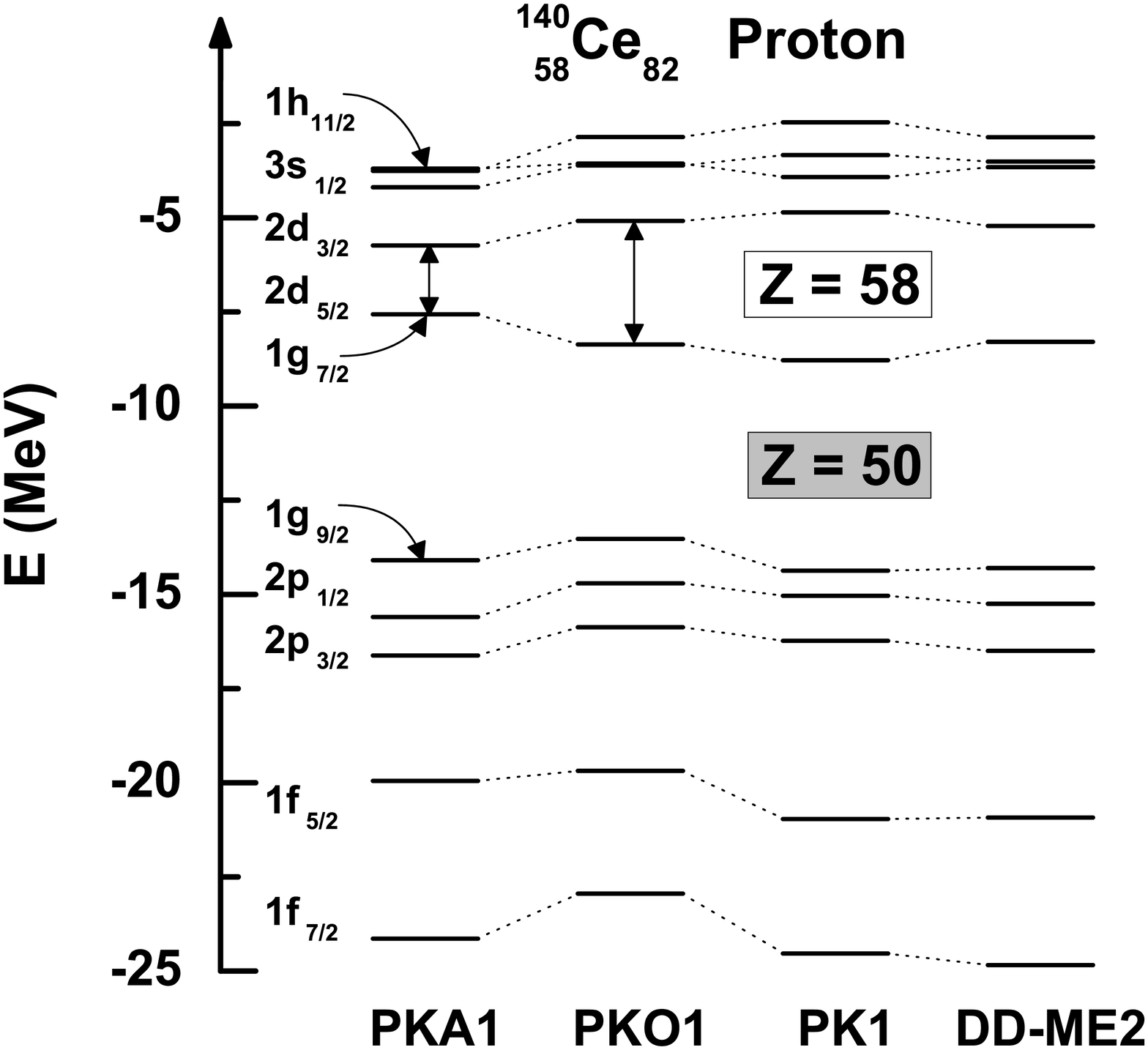}
\caption{Single-particle energies of $^{140}$Ce. The results are calculated by DDRHF with PKA1 and PKO1, and RMF with
PK1 and DD-ME2. }\label{fig:Ce140}
 \end{figure}

\begin{figure}[htbp]
\includegraphics[width = 0.40\textwidth]{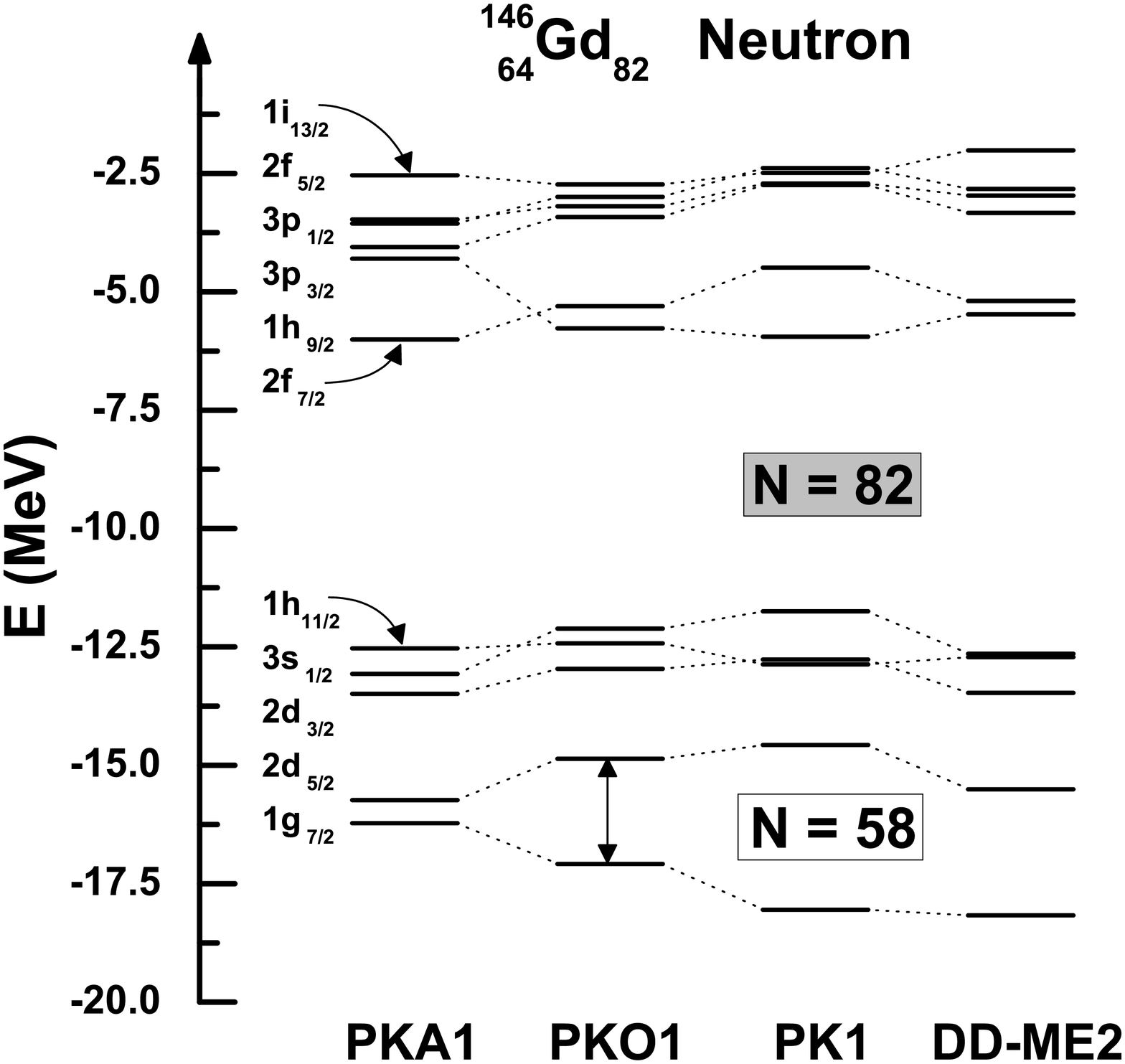}
\includegraphics[width = 0.40\textwidth]{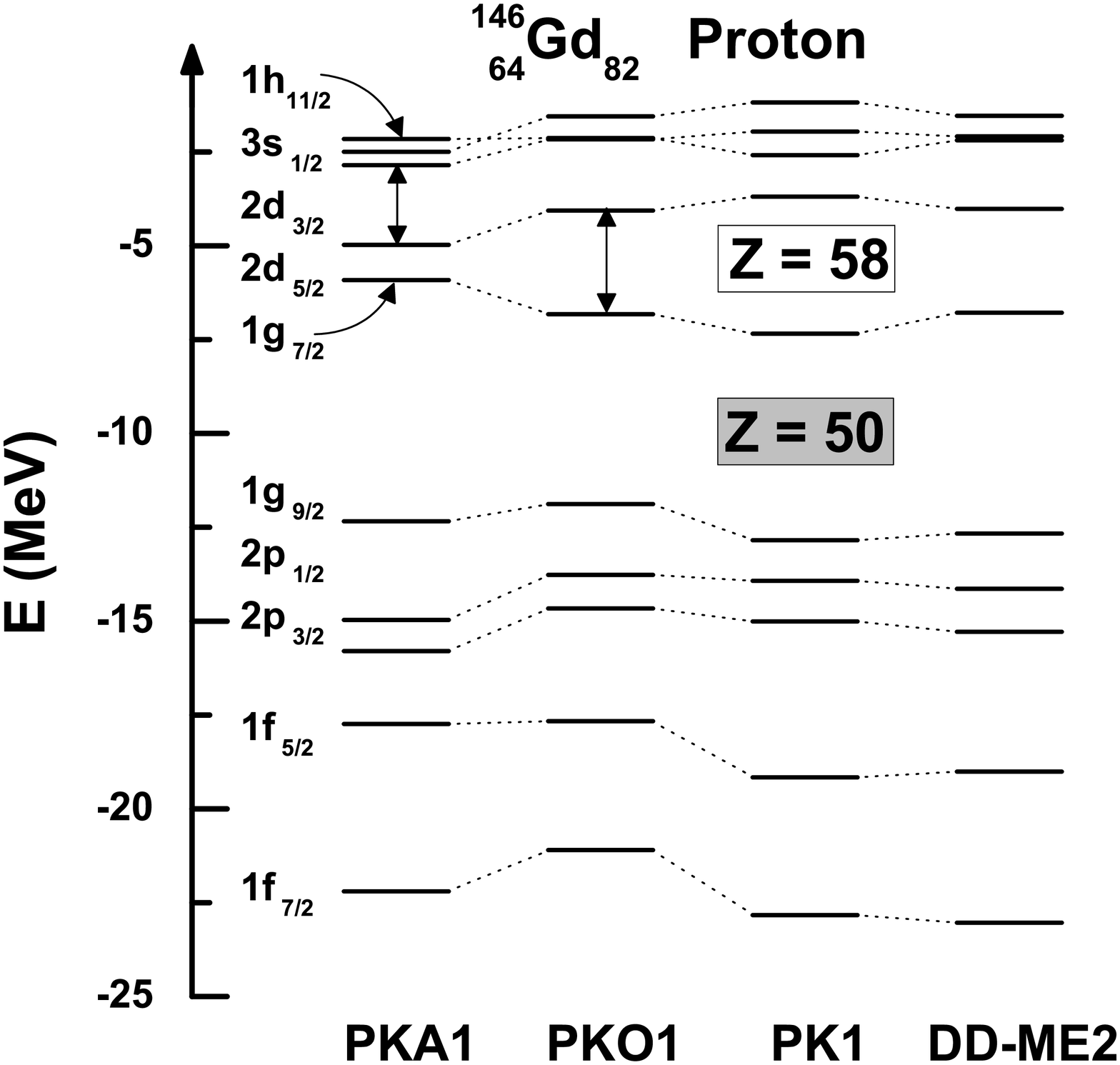}
\caption{Single-particle energies of $^{146}$Gd. The results are calculated by DDRHF with PKA1 and PKO1, and RMF with
PK1 and DD-ME2. }\label{fig:Gd146}
 \end{figure}

In \figref{fig:Ce140} and \figref{fig:Gd146} are shown the neutron (left panel) and proton (right panel)
single-particle levels in $^{140}$Ce and $^{146}$Gd, respectively. For these two $N=82$ isotones, their proton numbers
correspond to the artificial shell closure $Z=58$ occurring in RMF and the sub-shell closure $Z=64$ observed in
experiments, respectively. In the results calculated with PKO1, PK1, and DD-ME2, one can find large shell gaps at $Z
=58$ or $N = 58$ in both nuclei. These gaps are {even} comparable to the well established shell gaps $Z = 50$ and $N =
82$. In contrast, PKA1 brings much smaller shell gaps at $N=58$ and $Z =58$ in the single-particle levels of $^{140}$Ce
In $^{146}$Gd, the artificial shell structures $Z=58$ and $N=58$ completely disappear and the sub-shell closures $Z =
64 $ and $N=64$ are well reproduced by PKA1.

\begin{figure}[htbp]
\includegraphics[width = 0.48\textwidth]{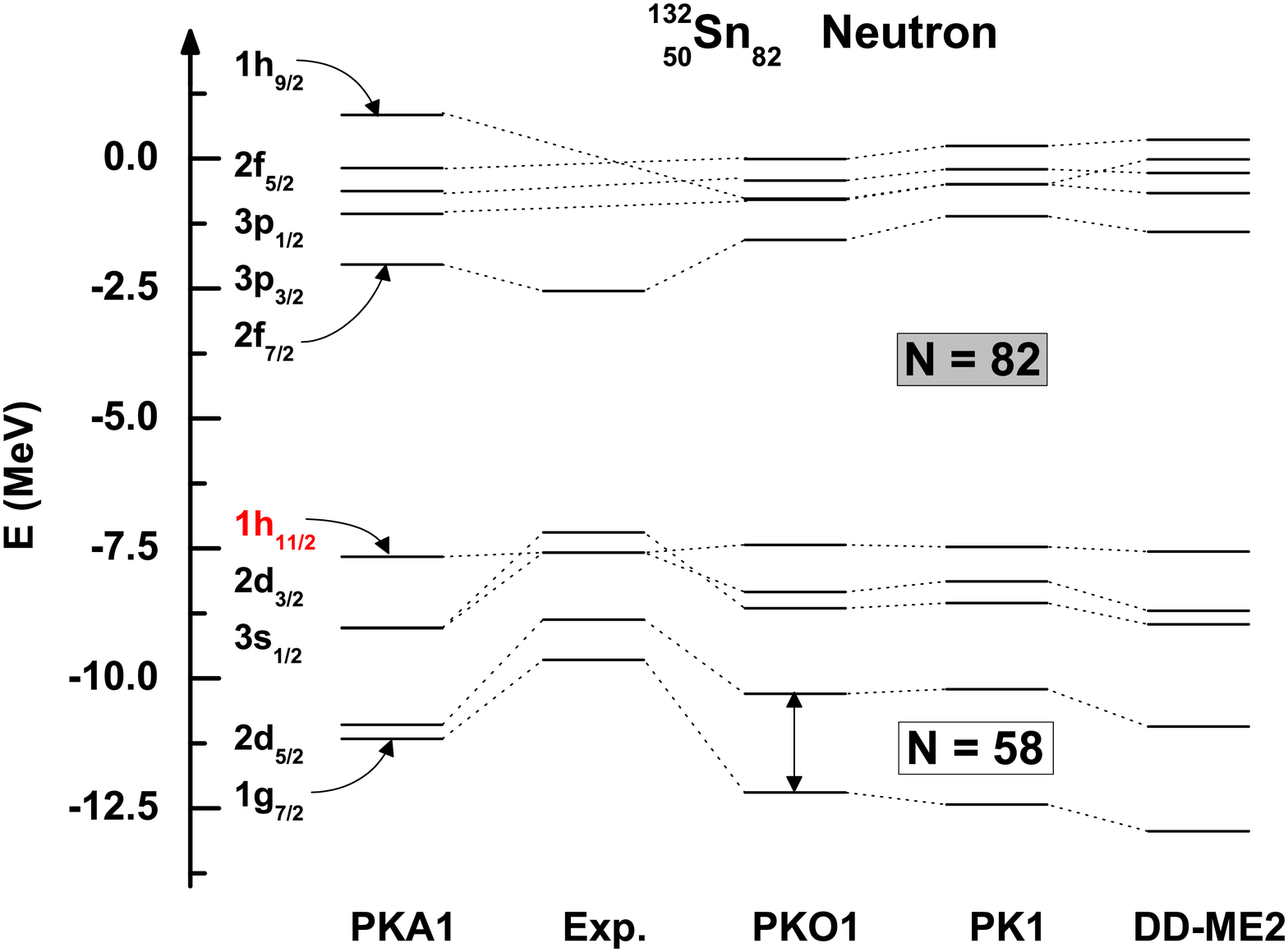}
\includegraphics[width = 0.48\textwidth]{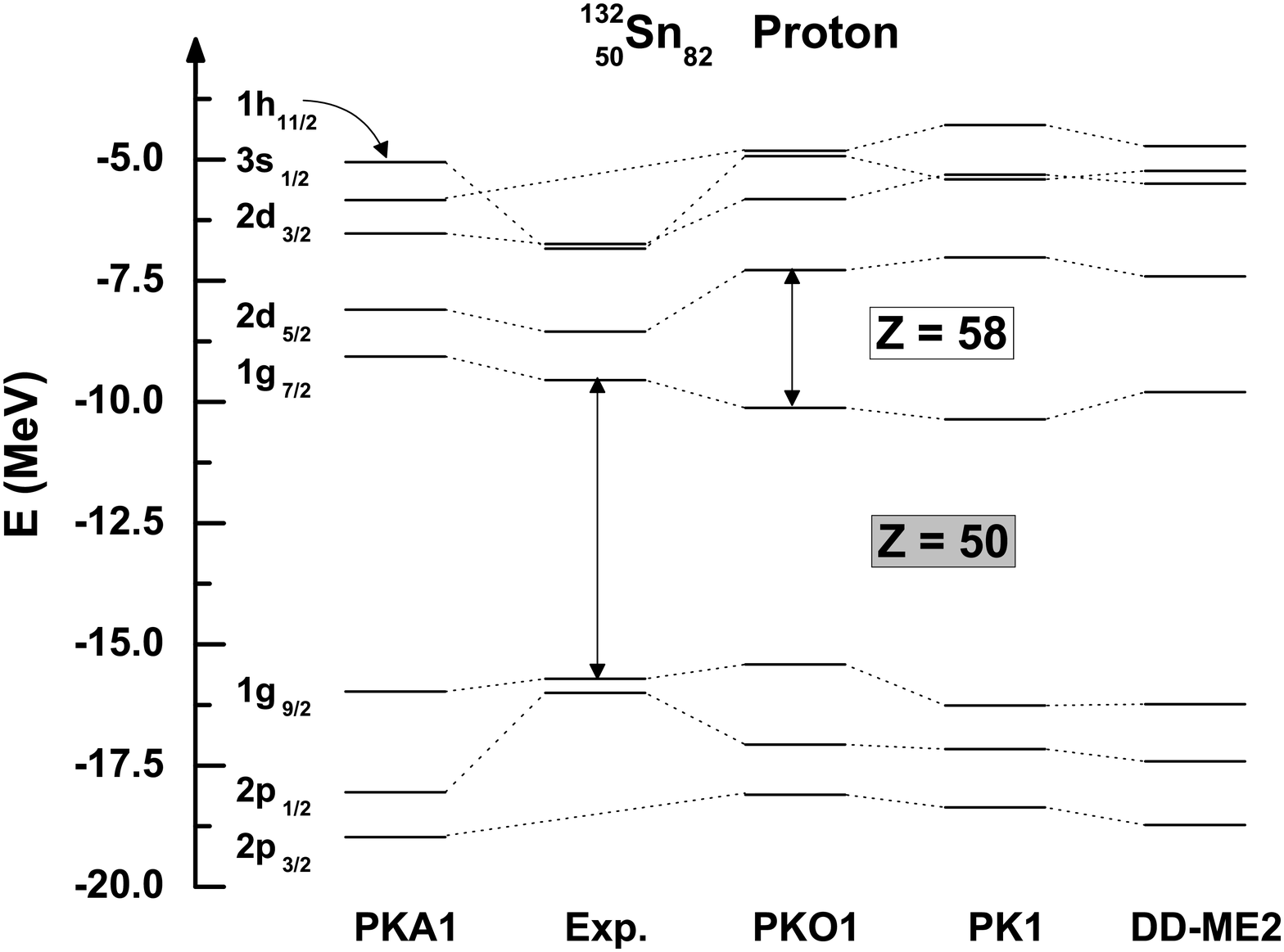}
\caption{Single-particle energies of $^{132}$Sn. The results are calculated by DDRHF with PKA1 and PKO1, and RMF with
PK1 and DD-ME2. Experimental data are taken from Ref. \cite{Oros:1996}.}\label{fig:Sn132}
 \end{figure}

Besides $^{140}$Ce and $^{146}$Gd, we also calculated another $N=82$ isotone, the doubly magic nucleus $^{132}$Sn. In
\figref{fig:Sn132} are shown the  neutron (left panel) and proton (right panel) single-particle levels calculated with
PKA1, PKO1, PK1, and DD-ME2. The experimental data from Ref. \cite{Oros:1996} are also shown for comparison. In this
figure, one can also find the spurious shell structures $Z$ or $N=58$ appearing in the results of PKO1, PK1, and
DD-ME2. Compared to these three results, PKA1 shows distinct improvement on this problem. Namely, the spurious shell
gaps do not appear any more and fairly good agreement with the data is obtained with PKA1.

\begin{figure}[htbp]
\includegraphics[width = 0.48\textwidth]{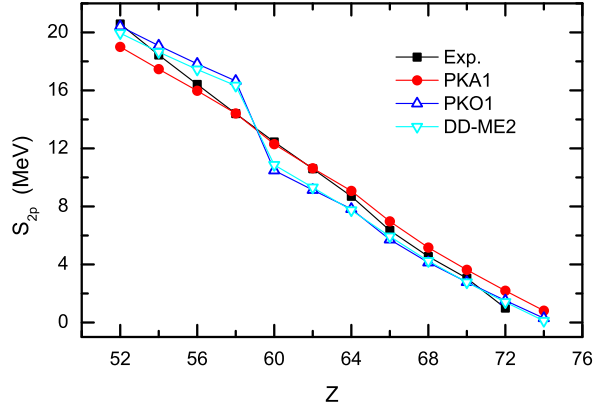}
\caption{(color online) Two-proton separation energies of the $N=82$ isotones. The results are calculated by DDRHF with
PKA1 and PKO1, and RMF with DD-ME2. The experimental data are taken from Ref. \cite{Audi:2003}.}\label{fig:S2pN82}
 \end{figure}

{It is well known that the two-nucleon  separation energy is an important criteria to identify the nuclear shell
structure. In \figref{fig:S2pN82}, the two-proton separation energy $S_{2p}$ calculated with PKA1, PKO1 and DD-ME2 are
shown as a function of proton number $Z$ for the $N=82$ isotones, and the experimental data \cite{Audi:2003} are also
shown for comparison. A sudden change of the slope is found in the results of} {PKO1 and DD-ME2 at $Z=58$. This is a
clear sign of the existence of spurious shell structure at $Z=58$ in the calculations of the existing DDRHF and RMF
Lagrangians. In contrast, the two-proton separation energy calculated by PKA1 decreases smoothly as a function of $Z$
and agrees well with the experimental data. \figref{fig:S2pN82} shows that the spurious shell structure $Z=58$ is
successfully eliminated by the new DDRHF effective interaction PKA1.}

\begin{figure}[htbp]
\includegraphics[width = 0.48\textwidth]{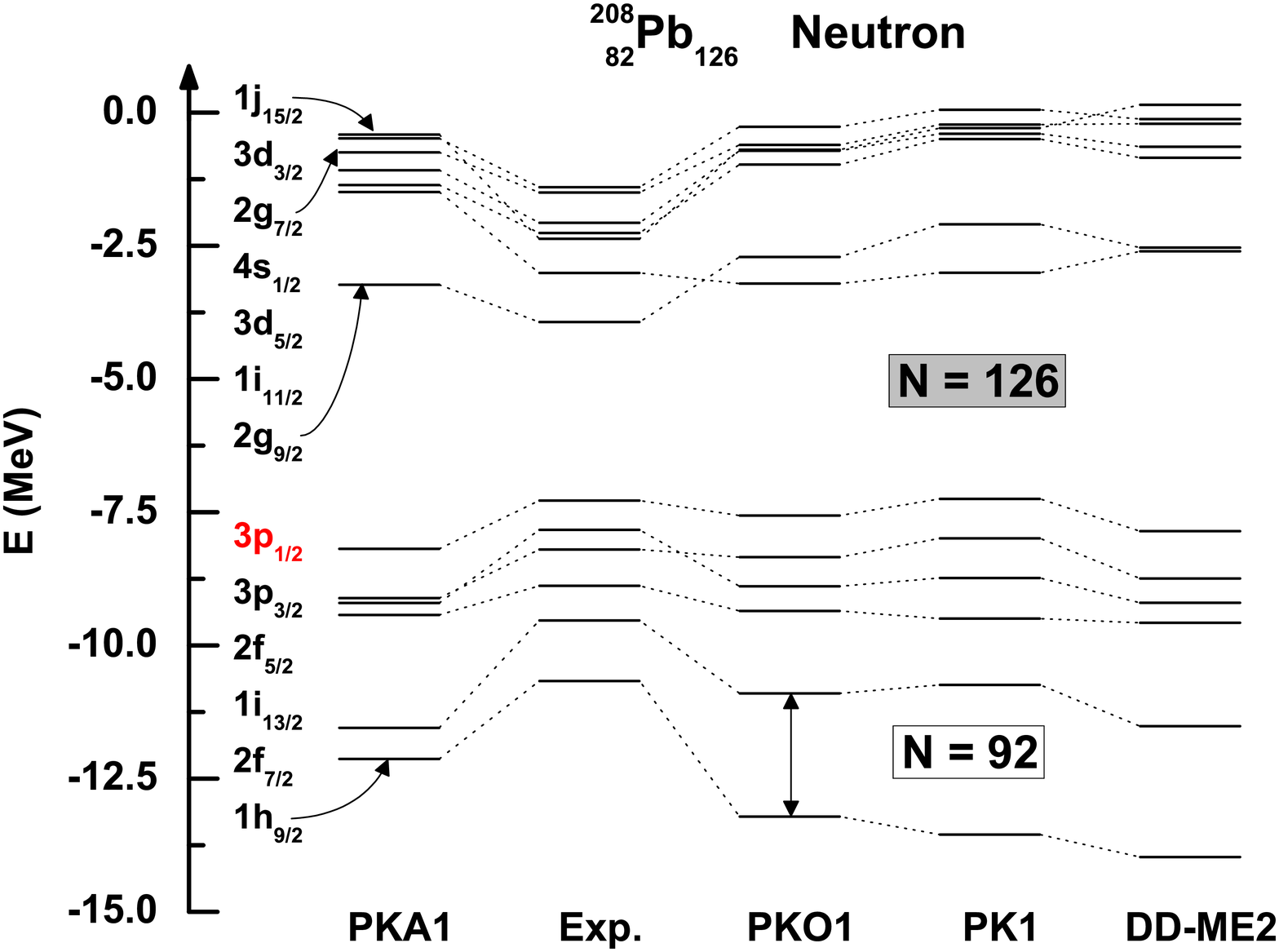}
\includegraphics[width = 0.48\textwidth]{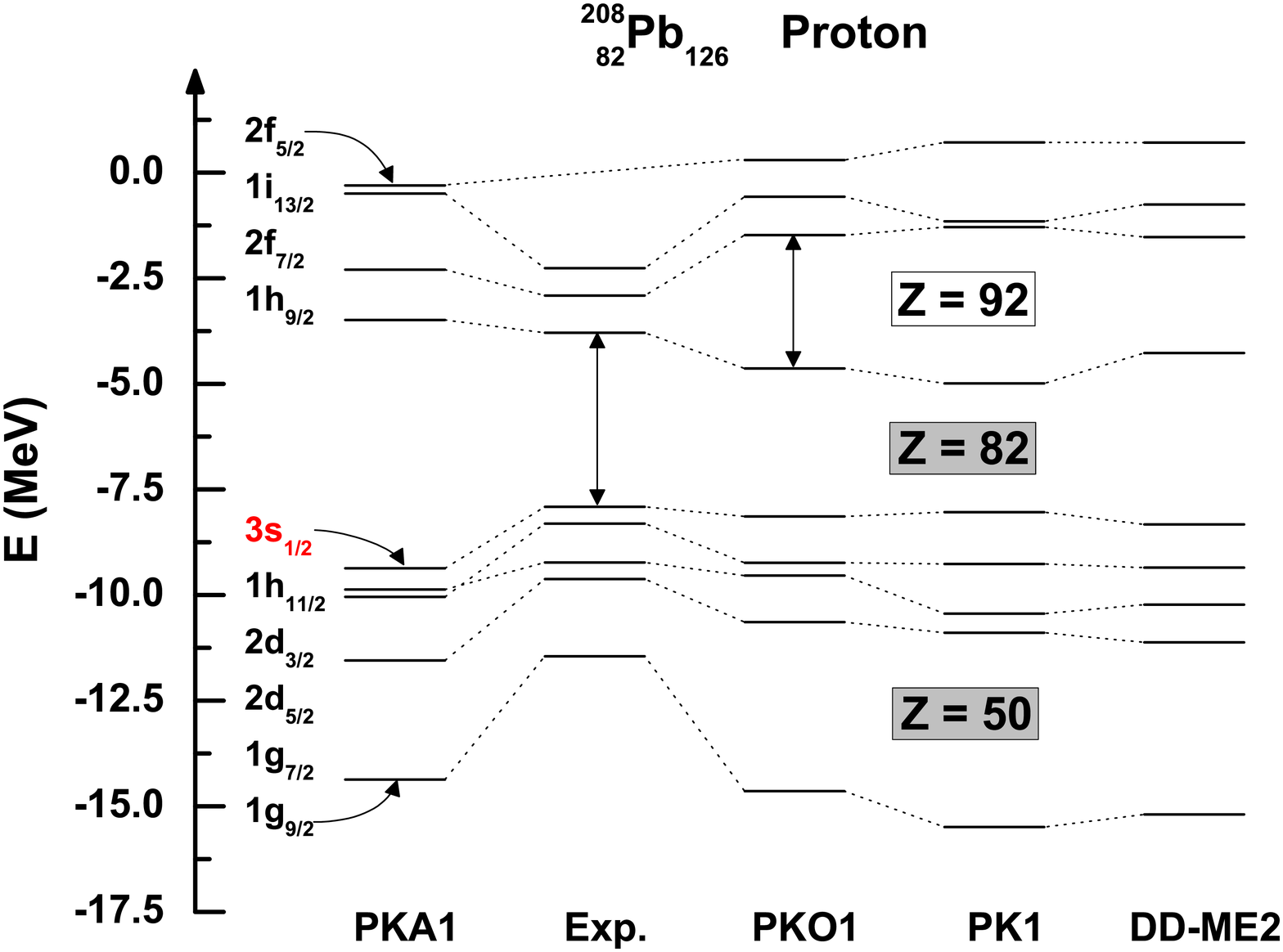}
\caption{Single-particle energies of $^{208}$Pb. The results are calculated by DDRHF with PKA1 and PKO1, and RMF with
PK1 and DD-ME2. The experimental data are taken from Ref. \cite{Oros:1996}.}\label{fig:Pb208}
 \end{figure}

Besides $N$ (or $Z$) $=58$, the nucleon number $92$ is another spurious shell structure in the calculations of RMF
\cite{Geng:2006}. In \figref{fig:Pb208}, it is clearly shown in both neutron and proton single-particle spectra that
$N$ (or $Z$) $=92$ becomes a spurious shell structure in the results of PKO1, PK1, and DD-ME2. With the inclusion of
the $\rho$-tensor correlations, the artificial shell closures at $N=92$ and $Z=92$ disappear in both neutron and proton
spectra in the results of PKA1. In addition, PKA1 also shows another improvement in the order of single-particle
levels, e.g., for the neutron states $2g_{9/2}$ and $1i_{11/2}$ in $^{208}$Pb. Namely, DDRHF with PKA1 provides the
same ordering with the data while the other three cases fail to predict the correct ordering.

\subsection{Spin-orbit Splittings and $\rho$-Tensor Correlations}

From the discussions on the single-particle spectra of several reference nuclei, one can find significant improvements
with the inclusion of $\rho$-tensor correlations in DDRHF calculations. The spin-orbit splitting is also very essential
to test the validity of the model. In \tabref{tab:splittings}, we give the spin-orbit splittings in several magic
nuclei obtained by using PKA1, PKO1, and DD-ME2. The experimental data are also tabulated for comparison. As seen from
this table, PKA1 provides comparable quantitative results of the spin-orbit splittings to PKO1 and DD-ME2. Among the
results calculated with PKA1, there are some systematic over estimations of the spin-orbit splittings of some states,
e.g., $\nu$1f in $^{48}$Ca and $^{56}$Ni, $\nu$1g in $^{90}$Zr, $\pi$1g in $^{132}$Sn, and $\nu$1i, $\pi$1h in
$^{208}$Pb, which account for the corresponding shell closures. For these spin partner states, one of the partners is
not occupied and one may expect sizable corrections from particle-vibration coupling. In the present calculations, the
particle-vibration coupling is not included yet, and its effects generally tend to shift the occupied and unoccupied
states to the Fermi surface. Therefore, the systematic over estimation of the spin-orbit splittings may leave some
space for this effect.

\begin{table}[htbp]\setlength{\tabcolsep}{6pt}
\caption{The spin-orbit splittings (in MeV) of neutron ($\nu$) and proton ($\pi$) states for the magic nuclei
calculated with PKA1, PKO1, and DD-ME2. The experimental data are taken from Ref. \cite{Oros:1996}.
}\label{tab:splittings}
\begin{tabular}{cccccc|cccrcc}\toprule[1.5pt]\toprule[0.5pt]
 Nucleus   & State   &Exp.  & PKA1  & PKO1  &DD-ME2 &  Nucleus   & State   & Exp. & PKA1  & PKO1  &DD-ME2 \\ \midrule[0.5pt]
 $^{16}$O  & $\nu$1p & 6.18 & 6.055 & 6.426 & 6.545 &  $^{56}$Ni & $\nu$1f & 6.82 &10.027 & 7.363 & 8.361 \\
           & $\pi$1p & 6.32 & 5.973 & 6.356 & 6.472 &            & $\nu$2p & 1.11 & 0.899 & 0.977 & 1.383 \\ \midrule[0.5pt]
 $^{40}$Ca & $\nu$1d & 6.75 & 7.386 & 6.742 & 6.760 &  $^{90}$Zr & $\nu$2p & 0.37 & 1.702 & 1.598 & 1.686 \\
           & $\nu$2p & 2.00 & 2.527 & 1.846 & 1.694 &            & $\nu$2d & 2.42 & 2.457 & 2.049 & 2.097 \\
           & $\pi$1d & 5.94 & 7.215 & 6.629 & 6.696 &            & $\nu$1g & 7.07 & 8.608 & 7.155 & 7.609 \\ \cmidrule[0.5pt]{1-6}
 $^{48}$Ca & $\nu$1d & 5.30 & 6.817 & 5.414 & 6.172 &            & $\pi$2p & 1.51 & 1.489 & 1.430 & 1.619 \\ \cmidrule[0.5pt]{7-12}
           & $\nu$1f & 8.01 & 8.512 & 7.345 & 7.737 &  $^{208}$Pb& $\nu$2f & 2.14 & 2.342 & 2.009 & 2.317 \\
           & $\nu$2p & 1.67 & 1.647 & 1.347 & 1.462 &            & $\nu$2g & 2.38 & 2.482 & 2.103 & 2.322 \\
           & $\pi$1d & 5.01 & 6.833 & 5.590 & 6.406 &            & $\nu$1i & 5.81 & 7.936 & 6.143 & 6.970 \\
           & $\pi$2p & 2.14 & 1.634 & 1.322 & 1.539 &            & $\nu$3p & 0.90 & 0.925 & 0.782 & 0.889 \\ \cmidrule[0.5pt]{1-6}
$^{132}$Sn & $\nu$2d & 1.66 & 1.866 & 1.645 & 1.969 &            & $\nu$3d & 0.89 & 0.875 & 0.710 & 0.726 \\
           & $\pi$1g & 5.33 & 6.909 & 5.291 & 6.438 &            & $\pi$2d & 1.53 & 1.506 & 1.404 & 1.769 \\
           & $\pi$2d & 1.75 & 1.569 & 1.463 & 1.912 &            & $\pi$1h & 5.03 & 6.380 & 4.901 & 5.955 \\
\bottomrule[0.5pt]\bottomrule[1.5pt]
\end{tabular}
\end{table}

From the Dirac equation, one can express the single-particle energy for a state $a$ as
 \beq \label{Single-particle}
E_a = E_{k, a} + E_{\sigma, a} + E_{\omega, a} + E_{\rho, a} + E_{\pi, a} + E_{A, a} + E_{R, a}~,
 \eeq
where $E_{k, a}$ denotes the kinetic contribution, and $E_{i,a}$ ($i = \sigma, \omega, \rho, \pi, A$) represent the
contributions from the mesons and photon coupling channels including the direct and exchange parts, and $E_{R,a}$
accounts for the rearrangement terms. From Eq. (\ref{Single-particle}), one can also obtain the contributions to the
spin-orbit splittings from different channels.

 \begin{figure}[htbp]
\includegraphics[width = 0.48\textwidth]{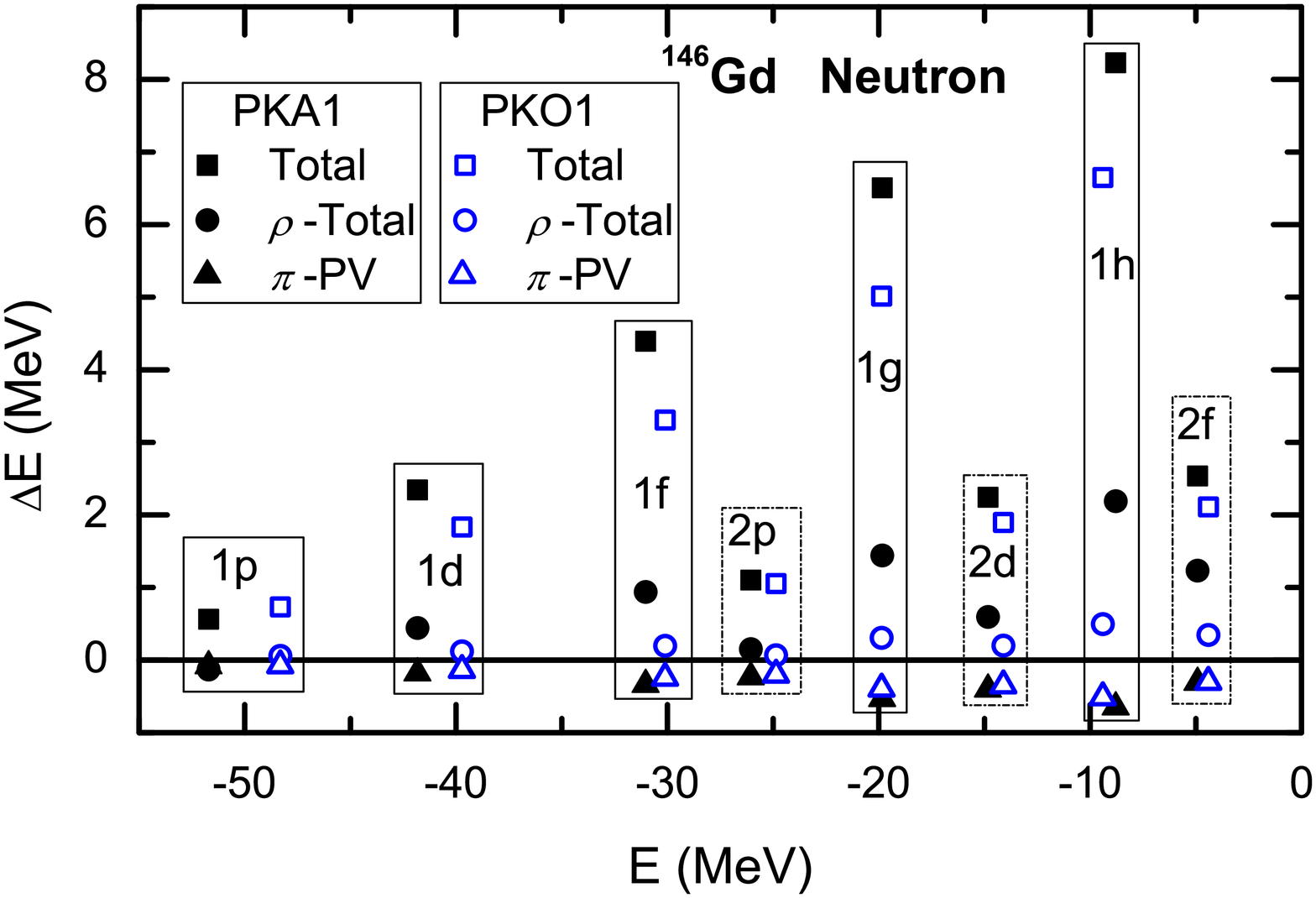}
\includegraphics[width = 0.48\textwidth]{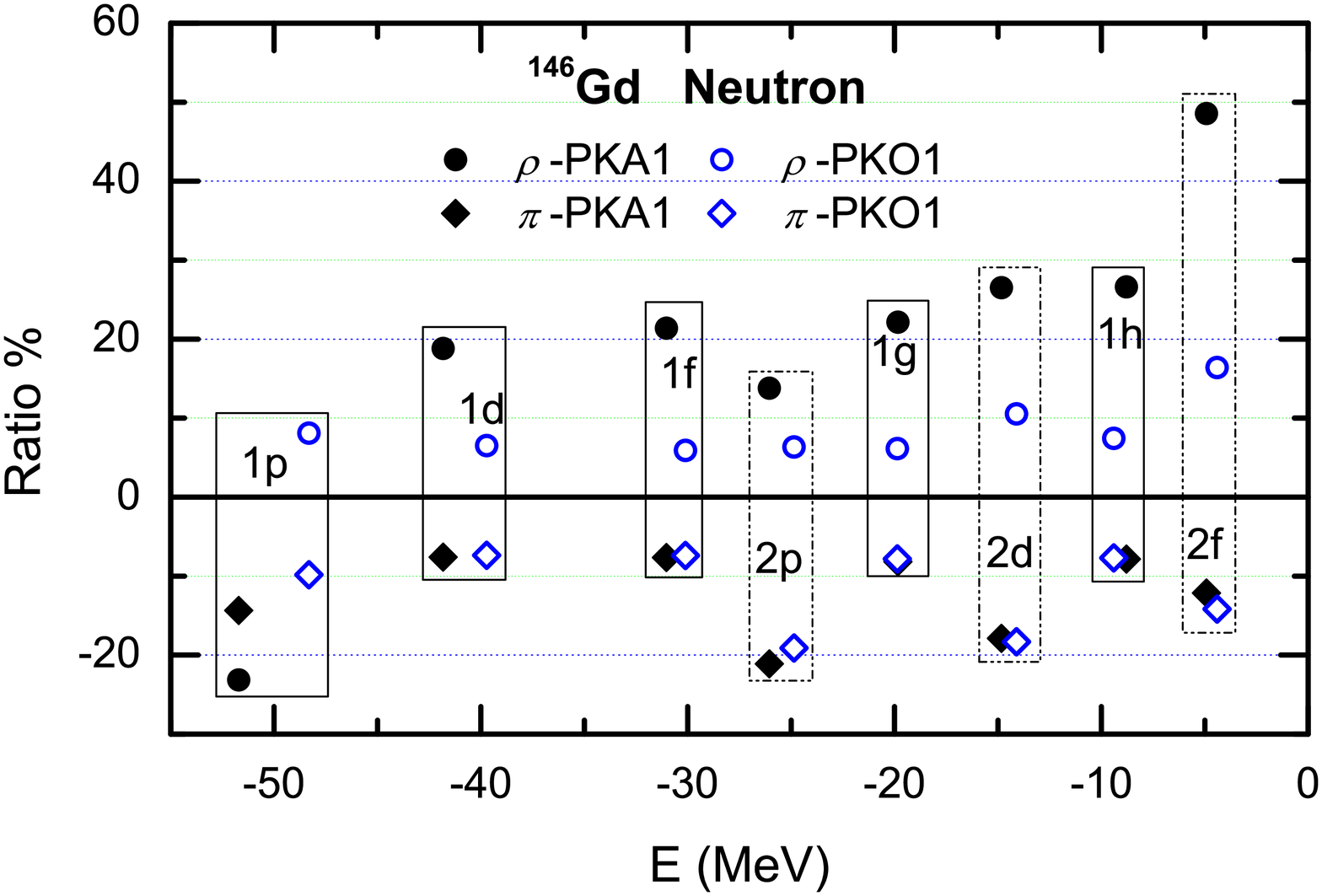}
\caption{Contributions from $\rho$- and $\pi$-mesons to the spin-orbit splittings for the neutron orbits in $^{146}$Gd
calculated by DDRHF with PKA1 and PKO1.The left panel shows the total values of the spin-orbit splittings and the
contributions from the $\rho$- and $\pi$-mesons, and the right panel shows the ratios of the $\rho$- and $\pi$-mesons
contributions to the spin-orbit splittings. }\label{fig:Gd146Spin}
 \end{figure}

 \begin{figure}[htbp]
\includegraphics[width = 0.48\textwidth]{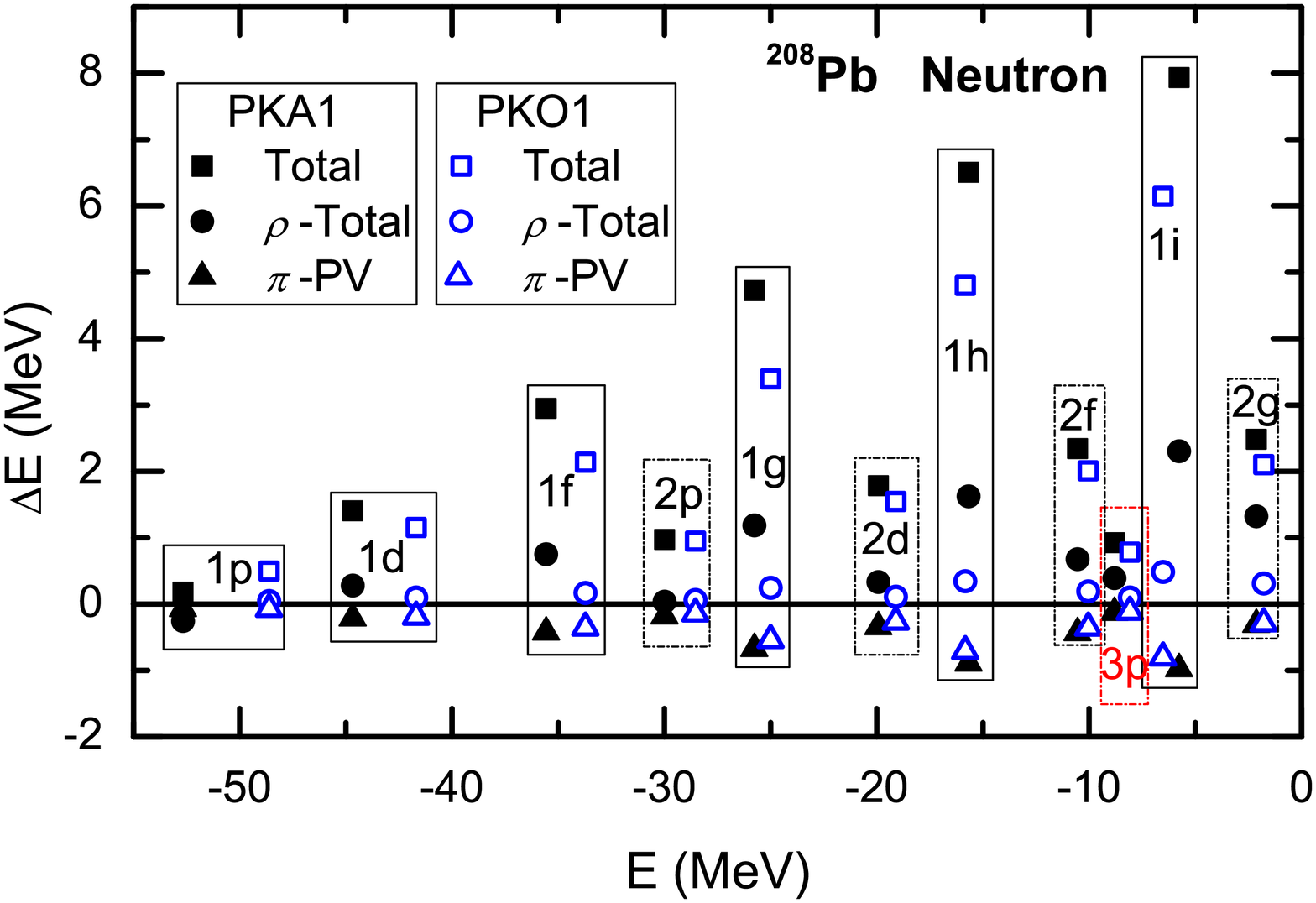}
\includegraphics[width = 0.48\textwidth]{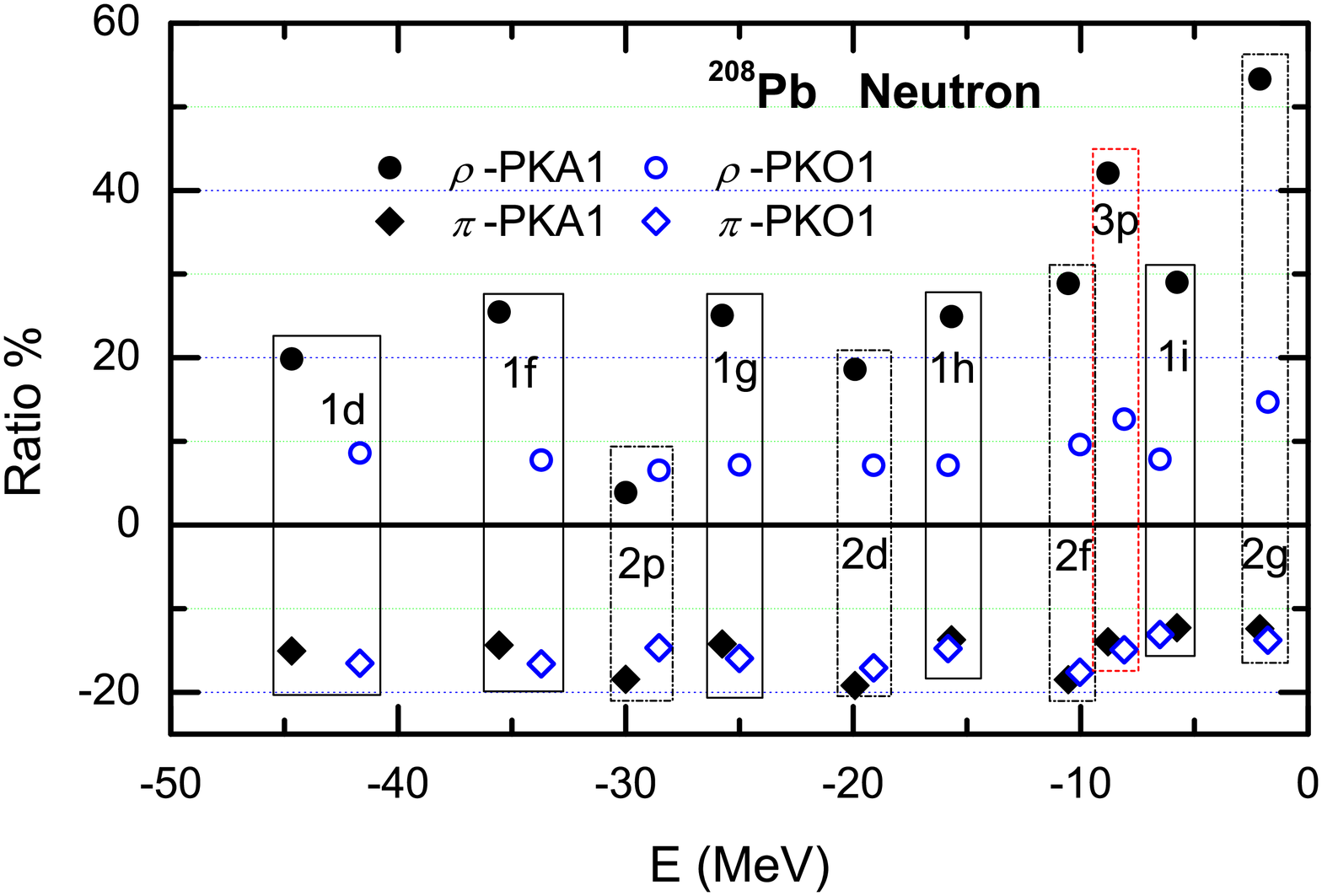}
\caption{Similar as \figref{fig:Gd146Spin}, but for $^{208}$Pb. }\label{fig:Pb208Spin}
 \end{figure}

In order to understand the improvement on the shell structure brought by the $\rho$-tensor couplings, we compare the
contributions of the $\rho$- and $\pi$-mesons to the spin-orbit splittings between two DDRHF effective interactions
PKA1 and PKO1. In \figref{fig:Gd146Spin} and \figref{fig:Pb208Spin}, are shown the contributions to the spin-orbit
splittings from $\rho$- and $\pi$-mesons, respectively for the $^{146}$Gd and $^{208}$Pb neutron orbits. From these two
figures, one can find that the $\pi$ pseudo-vector (PV) coupling gives almost same contributions to the spin-orbit
splittings in PKA1 (filled up-triangles) and PKO1 (open up-triangles). This can be well explained by the equivalent
$\pi$-coupling strength in these two effective interactions. For the contributions from $\rho$-meson, PKA1 and PKO1
have distinct difference. Due to the tensor interactions, the $\rho$-meson coupling in PKA1 (filled circles) shows
substantially larger effects than that of PKO1 (open circles). It is seen that the average contributions from the
$\rho$-meson couplings in PKA1 are about 20\% of the total ones. For the states near the Fermi surface or some high-$j$
orbits, the $\rho$-meson contribution grows up to 30\% and even higher.  On the other hand, the $\rho$-meson coupling
of PKO1 gives about 10\% in most cases. The $\pi$-meson gives opposite contributions to the $\rho$-meson for the
spin-orbit splittings except the $1p$ states and the magnitude is about (10$\sim$ 20)\% of the spin-orbit splitting.
Because of the tensor effects in $\pi$-PV couplings, some systematic enhancements are also observed in high-$j$ states
like $1g$ and $1h$ states in $^{146}$Gd, and ${1h}$ and $1i$ in $^{208}$Pb. As was noticed in the comparison between
PKA1 and PKO1, the $\rho$-tensor couplings have significant effects on the spin-orbit splittings, especially for
high-$j$ orbits, and affect much the shell structures. {This is the main reason why the improvement of the shell
structure is obtained with the $\rho$-tensor correlations.}

\subsection{Spurious Shell closures and Pseudo-spin symmetry}

As we mentioned before, the spurious shell closures at 58 and 92 are related with the pairs of high-$j$ states
$\Lrb{2d_{5/2}, 1g_{7/2}}$ and $\Lrb{2f_{7/2}, 1h_{9/2}}$, respectively. These pairs are the pseudo-spin partners,
$1\tilde f$ and $1\tilde g$ states, respectively. The spurious shell closure problem is then related to the
conservation of pseudo-spin symmetry (PSS) \cite{Arima:1969, Hecht:1969, Meng:1998PRCps, Meng:1999PRC, ChenTS:2003,
ginocchio:2005, Long:2006PS}, i.e., the existing artificial shell structures in RMF break largely PSS. As seen from the
results of PKA1 (see \figref{fig:Gd146} and \figref{fig:Pb208}), PSS is successfully recovered for the $1\tilde f$
states in $^{146}$Gd and $1\tilde g$ states in $^{208}$Pb. In order to understand the improvement due to the
$\rho$-tensor correlations, we studied the contributions from different terms in Eq. (\ref{Single-particle}) to the
pseudo-spin orbital splittings. In \tabref{tab:pseudo-Gd146} and \tabref{tab:pseudo-Pb208} are shown the results
calculated by DDRHF with PKA1 (upper panels) and PKO1 (lower panels) respectively for $^{146}$Gd and $^{208}$Pb neutron
orbits. One can find in these results that PKA1 conserves pseudo-spin symmetry better than PKO1 for the states near the
Fermi levels, e.g., $\nu1\tilde f$ and $\nu2\tilde p$ states in $^{146}$Gd, $\nu1\tilde g$ and $\nu2\tilde d$ states in
$^{208}$Pb.

\begin{table}[htbp]\setlength{\tabcolsep}{3pt}
\caption{The contributions (in MeV) from different terms in Eq. (\ref{Single-particle}) to the pseudo-spin orbital
splittings $\Delta E$ for neutron ($\nu$) orbits in $^{146}$Gd , calculated by DDRHF with PKA1 (upper panel) and PKO1
(lower panel). The average binding energy $\bar E$ for the pseudo-spin partner states $j_1$ and $j_2$ is
$\lrs{E_1(2j_1+1)+E_2(2j_2+1)}/\lrb{2j_1+2j_2+2}$. }\label{tab:pseudo-Gd146}
\begin{tabular}{c|c|cc|crrcc|rr}\toprule[1.5pt]\toprule[0.5pt]
          $^{146}$Gd     &     State     &$\bar E$&$\Delta E$&$\Delta E_\rho$&$\Delta E_\pi$&$\Delta E_R$~~ &$\Delta E_k$&$\Delta E_{\sigma+\omega}$&$\Delta E_\sigma$&$\Delta E_\omega$\\ \midrule[0.5pt]
\multirow{4}{2.5em}{PKA1}&$\nu1\tilde p$ &-39.44  &  2.935   &  0.277        &        0.039 &   0.080       &   1.900    &  0.639                   &    11.391       &     -10.751     \\
                         &$\nu1\tilde d$ &-27.68  &  2.097   &  0.476        &        0.225 &  -0.439       &   2.592    & -0.756                   &    14.847       &     -15.603     \\
                         &$\nu1\tilde f$ &-16.01  &  0.489   &  1.107        &        0.525 &  -1.187       &   2.546    & -2.502                   &    18.145       &     -20.647     \\
                         &$\nu2\tilde p$ &-13.35  &  0.422   &  0.602        &        0.344 &  -0.456       &   0.854    & -0.920                   &     7.499       &      -8.419     \\ \midrule[0.5pt] \midrule[0.5pt]
\multirow{4}{2.5em}{PKO1}&$\nu1\tilde p$ &-37.63  &  2.985   &  0.429        &        0.041 &  -0.370       &   1.521    &  1.364                   &     8.502       &      -7.138     \\
                         &$\nu1\tilde d$ &-27.02  &  3.013   &  0.493        &        0.180 &  -0.814       &   2.131    &  1.023                   &    12.372       &     -11.349     \\
                         &$\nu1\tilde f$ &-16.13  &  2.224   &  0.662        &        0.404 &  -1.360       &   1.945    &  0.573                   &    18.306       &     -17.733     \\
                         &$\nu2\tilde p$ &-12.68  &  0.851   &  0.387        &        0.284 &  -0.524       &   0.188    &  0.516                   &     9.315       &      -8.799     \\
\bottomrule[0.5pt]\bottomrule[1.5pt]
\end{tabular}
\end{table}

For the pseudo-spin orbital splittings, PKA1 and PKO1 provide similar contributions in magnitude to the kinetic part
($\Delta E_k$), the rearrangement term ($\Delta E_R$), and the $\pi$-coupling ($\Delta E_\pi$) except for a few cases.
For the contributions from $\sigma$-, $\omega$-mesons ($\Delta E_{\sigma+\omega}$) and $\rho$-meson ($\Delta E_\rho$)
couplings, there exist a distinct difference between PKA1 and PKO1, especially for the states near the Fermi surfaces.
From these two tables one can see that the $\rho$-meson couplings in PKA1 give larger contributions to the pseudo-spin
orbital splittings than {those in} PKO1 and the $\rho$-tensor couplings increase the splittings. For the states near
the Fermi surface, PKA1 provides negative values of $\Delta E_{\sigma+\omega}$, which cancel largely with $\Delta E_k$
and $\Delta E_\rho$. In the PKO1 results, the $\Delta E_{\sigma+\omega}$ is always positive and only the rearrangement
term $\Delta E_R$ partially cancels the contributions from the other channels.

\begin{table}[htbp]\setlength{\tabcolsep}{3pt}
\caption{Same as \tabref{tab:pseudo-Gd146}, for $^{208}$Pb neutron
orbits. }\label{tab:pseudo-Pb208}
\begin{tabular}{c|c|cc|crrcc|rr}\toprule[1.5pt]\toprule[0.5pt]
          $^{208}$Pb     &  State        &$\bar E$ &$\Delta E$&$\Delta E_\rho$&$\Delta E_\pi$&$\Delta E_R$~~ &$\Delta E_k$&$\Delta E_{\sigma+\omega}$&$\Delta E_\sigma$&$\Delta E_\omega$\\ \midrule[0.5pt]
\multirow{6}{2.5em}{PKA1}&$\nu1\tilde p$ &-42.74   &  3.292   &  0.342        &       0.083  &  -0.006       &   1.655    &  1.218                   &     7.555       &     -6.337 \\
                         &$\nu1\tilde d$ &-32.46   &  3.560   &  0.329        &       0.218  &  -0.111       &   2.393    &  0.730                   &    14.699       &    -13.969 \\
                         &$\nu1\tilde f$ &-22.06   &  2.502   &  0.608        &       0.455  &  -0.563       &   2.738    & -0.735                   &    20.507       &    -21.242 \\
                         &$\nu1\tilde g$ &-11.87   &  0.584   &  1.181        &       0.757  &  -1.251       &   2.431    & -2.535                   &    23.240       &    -25.776 \\
                         &$\nu2\tilde p$ &-18.76   &  0.259   &  0.265        &       0.233  &  -0.399       &   1.177    & -1.017                   &     3.704       &     -4.721 \\
                         &$\nu2\tilde d$ & -9.17   &  0.092   &  0.777        &       0.459  &  -0.695       &   0.773    & -1.222                   &     5.289       &     -6.510 \\ \midrule[0.5pt] \midrule[0.5pt]
\multirow{6}{2.5em}{PKO1}&$\nu1\tilde p$ &-40.04   &  2.851   &  0.238        &       0.100  &  -0.349       &   1.370    &  1.492                   &     4.663       &     -3.172 \\
                         &$\nu1\tilde d$ &-31.04   &  3.644   &  0.421        &       0.207  &  -0.712       &   2.009    &  1.720                   &    10.406       &     -8.686 \\
                         &$\nu1\tilde f$ &-21.66   &  3.397   &  0.549        &       0.380  &  -1.125       &   2.212    &  1.381                   &    16.130       &    -14.749 \\
                         &$\nu1\tilde g$ &-12.18   &  2.314   &  0.619        &       0.604  &  -1.530       &   1.731    &  0.890                   &    21.939       &    -21.048 \\
                         &$\nu2\tilde p$ &-17.95   &  0.678   &  0.117        &       0.181  &  -0.348       &   0.624    &  0.105                   &     4.354       &     -4.250 \\
                         &$\nu2\tilde d$ & -8.67   &  0.547   &  0.183        &       0.371  &  -0.566       &   0.124    &  0.434                   &     8.938       &     -8.503 \\
\bottomrule[0.5pt]\bottomrule[1.5pt]
\end{tabular}
\end{table}

\begin{figure}[htbp]
\includegraphics[width = 0.49\textwidth]{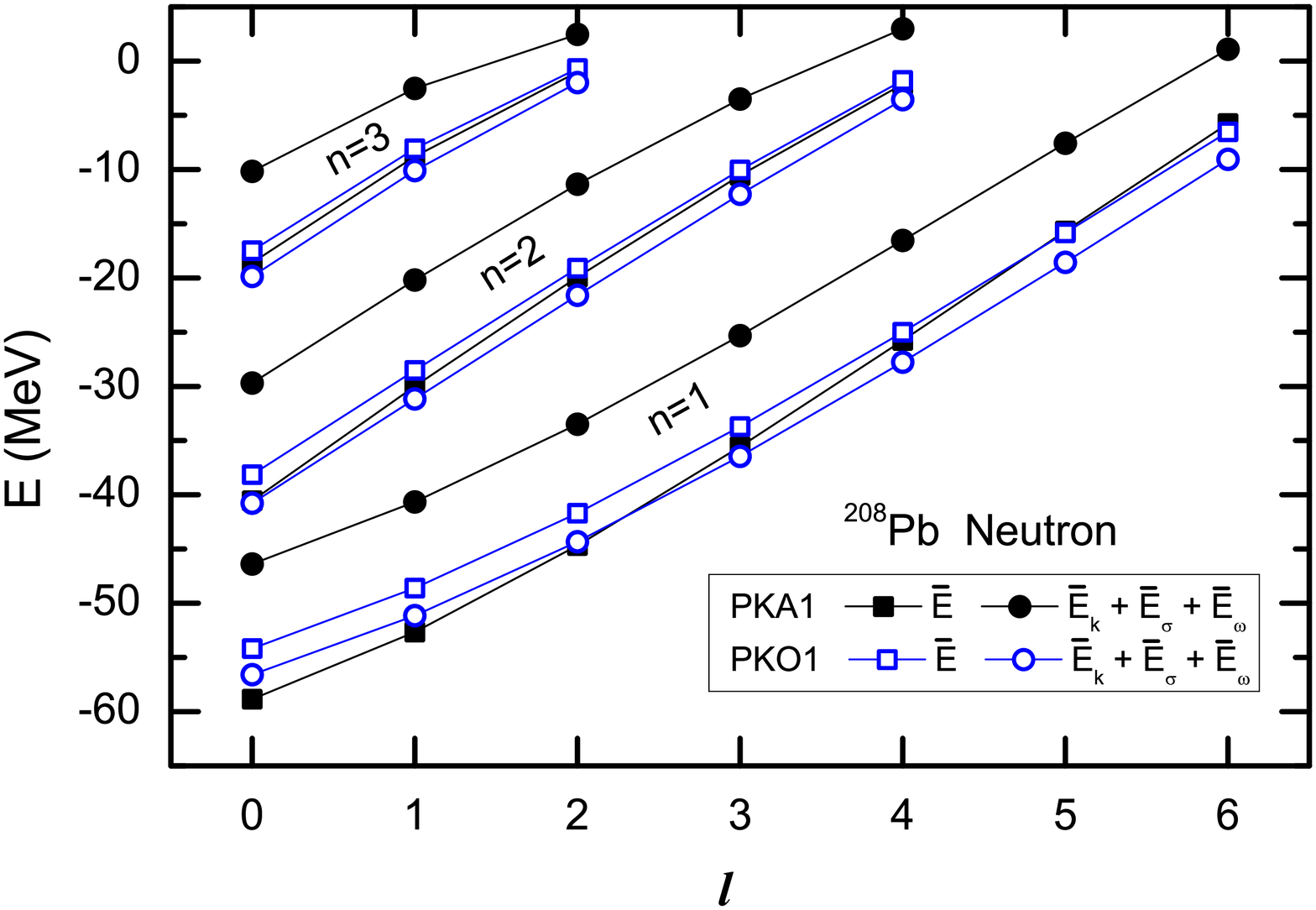}
\includegraphics[width = 0.49\textwidth]{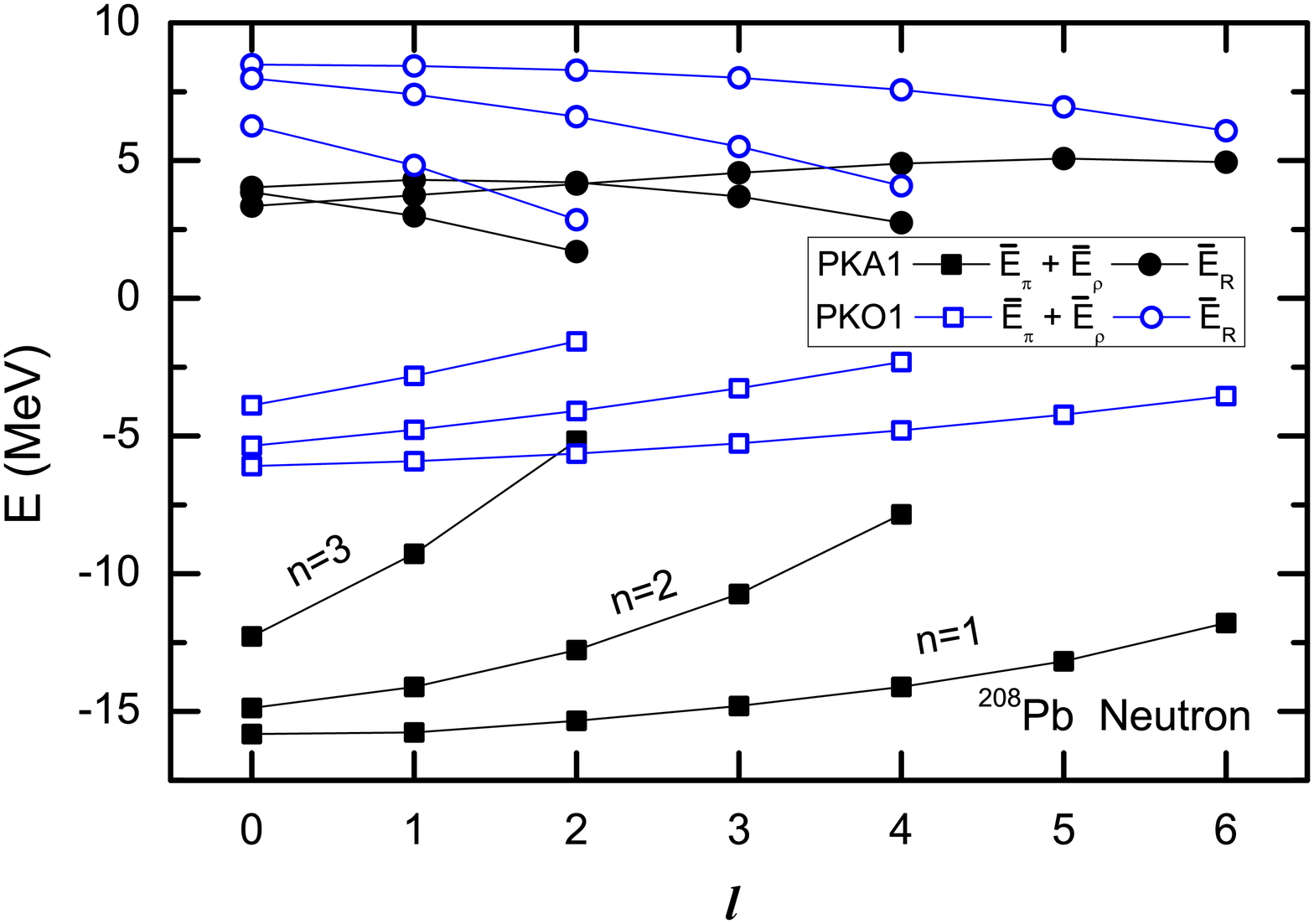}
\caption{The average binding energy $\bar E = \lrs{E_1(2j_1+1) + E_2(2j_2+1)}/\lrb{2j_1+2j_2+2}$ of the $^{208}$Pb
neutron orbits as a function of the angular momentum $l$ calculated by DDRHF with PKA1 (filled symbols) and PKO1 (open
symbols). Left panel gives the sum contributions from the kinetic part ($\bar E_k$), the $\sigma$- and
$\omega$-couplings ($\bar E_\sigma$ and $\bar E_\omega$) to the average binding energy and the total ones, and right
panel shows the contributions from the rearrangement term ($\bar E_R$), the $\rho$-, and $\pi$-couplings ($\bar E_\rho$
and $\bar E_\pi$). }\label{fig:NeutronNT}
 \end{figure}

To understand the differences between the results of PKA1 and PKO1, we study the contributions from different terms to
the average binding energy $\bar E$ of the spin partner states $j_1$ and $j_2$, i.e., $\bar E = \lrs{E_1(2j_1+1) +
E_2(2j_2+1)}/\lrb{2j_1+2j_2+2}$. In the left panel of \figref{fig:NeutronNT}, the average binding energies $\bar E$ and
the sum of the kinetic part ($\bar E_k$), the $\sigma$- and $\omega$-couplings ($\bar E_\sigma$ and $\bar E_\omega$)
are shown as a function of angular momentum $l$ for the neutron orbits of $^{208}$Pb. The right panel shows the
rearrangement term ($\bar E_R$), and the sum of the $\rho$- and $\pi$-coupling terms ($\bar E_\rho$ and $\bar E_\pi$).
The results are calculated with PKA1 (filled symbols) and PKO1 (open symbols). In the left panel, it is found that PKA1
provides stronger $l$-dependence than PKO1 for the average binding energy $\bar E$. In the results of PKO1, the main
contribution to $\bar E$ is given by the sum $\bar E_k + \bar E_\sigma + \bar E_\omega$, whereas in the results of
PKA1, $\bar E_R$ and $\bar E_\rho + \bar E_\pi$ also provide significant contributions to $\bar E$. This is due to the
fact that $\bar E_R$ and $\bar E_\rho + \bar E_\pi$ cancel each other in the results of PKO1 as shown in the right
panel. On the other hand, PKA1 gives weaker rearrangement term $\bar E_R$, and much stronger $\rho$-couplings than
PKO1. Thus, the inclusion of $\rho$-tensor couplings give significant contributions to the nuclear attraction, which
strongly affects on the coupling strength in other channels, e.g., PKA1 has a stronger $\omega$-coupling than PKO1 as
shown in \figref{fig:couplings}.

\begin{figure}[htbp]
\includegraphics[width = 0.49\textwidth]{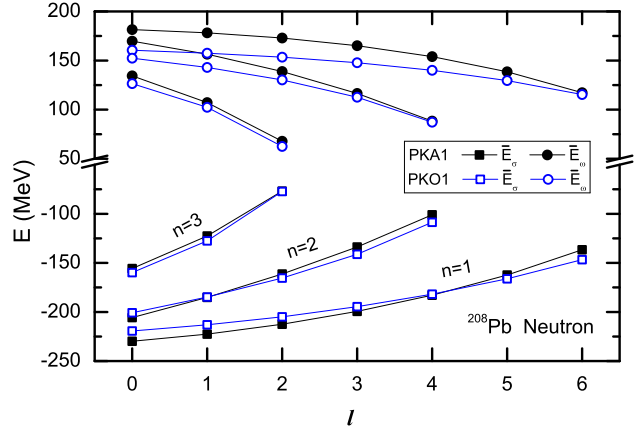}
\caption{Contributions from the $\sigma$- and $\omega$-couplings ($\bar E_\sigma$ and $\bar E_\omega$) to the average
binding energy $\bar E = \lrs{E_1(2j_1+1) + E_2(2j_2+1)}/\lrb{2j_1+2j_2+2}$ of the $^{208}$Pb neutron orbits as a
function of the angular momentum $l$ calculated with PKA1 (filled symbols) and PKO1 (open symbols).
}\label{fig:NeutronSW}
 \end{figure}

In \figref{fig:NeutronSW} are shown the values of $\bar E_\sigma$ and $\bar E_\omega$ calculated with PKA1 (filled
symbols) and PKO1 (open symbols) as a function of angular momentum $l$. One can see that PKA1 leads to a stronger
$l$-dependence of $\bar E_\omega$ than PKO1, and to a similar (slightly stronger) $l$-dependence of $\bar E_\sigma$ for
the states near the Fermi surface. It should be noticed that the pseudo-spin partner states have different angular
momenta $l$. The stronger $l$-dependence of $\bar E_\omega$ given by PKA1 leads to larger negative contributions to the
pseudo-spin orbital splittings as shown in the last column of \tabref{tab:pseudo-Gd146} and \tabref{tab:pseudo-Pb208}.
These results finally induce negative values for $\Delta E_{\sigma+\omega}$ so that PSS can be well conserved in the
results of PKA1. As a relativistic symmetry, the conservation of the PSS is mainly determined by the balance of the
nuclear attractions and repulsions \cite{Ginocchio:1997}, which is also well demonstrated by \tabref{tab:pseudo-Gd146}
and \tabref{tab:pseudo-Pb208}. Compared to the PKO1 results, this balance is much changed by PKA1 due to the extra
binding induced by the $\rho$-tensor couplings, which indicates the physical reason for the improvement of the nuclear
shell structure.

\section{Conclusions}\label{sec:conc}
In this work, we have introduced the $\rho$-tensor correlations in the density-dependent relativistic Hartree-Fock
(DDRHF) theory. By fitting the empirical properties of ground state and the shell structure, we propose a new DDRHF
effective interaction with $\rho$-tensor couplings, PKA1. With the newly obtained effective interaction PKA1, DDRHF
provides satisfactory descriptions of the bulk properties of nuclear matter and the ground state properties of finite
nuclei, at the same quantitative level as the established DDRHF and RMF models.

Moreover, the inclusion of $\rho$-tensor correlations brings a significant improvement on the descriptions of nuclear
shell structures compared to the existing DDRHF and RMF Lagrangians. Particularly, we have studied the single-particle
spectra of nuclei $^{140}$Ce, $^{146}$Gd, $^{132}$Sn and $^{208}$Pb with PKA1 and compared to previous DDRHF and RMF
approaches. It has been found that the previous DDRHF and RMF calculations give the spurious shell closures $58$ and
$92$, whereas the realistic sub-shell closure $64$ cannot be well reproduced. The effective interaction PKA1 cures
these common diseases, eliminating the spurious shell structure and recovering the sub-shell closure $64$. In addition,
the inclusion of tensor correlations improves the descriptions of the ordering of the single-particle levels, e.g., the
neutron states $2g_{9/2}$ and $1i_{11/2}$ in $^{208}$Pb, which are important states for nuclear structure problems.

The spin-orbit splittings and the pseudo-spin orbital splittings of the magic nuclei are also studied by using PKA1,
and the PKO1 version which has no $\rho$-tensor coupling. It is shown that the $\rho$-tensor correlations have
substantial effects on enlarging both splittings, especially for the high-$j$ states. Even though, PKA1 still provides
an appropriate quantitative agreement with the experimental data on the spin-orbit splittings for the magic nuclei at
the same level as the modern DDRHF and RMF Lagrangians. It is shown that the artificial shell structure problem is
intimately related to the conservation of the pseudo-spin symmetry, which is determined by the balance of the nuclear
attractions from $\sigma$-meson and $\rho$-tensor couplings and the repulsion from $\omega$-meson coupling. It is found
that a better conserved pseudo-spin symmetry is obtained with PKA1, in which the $\rho$-tensor correlations contribute
significantly to the nuclear attraction. Due to the extra binding introduced by the $\rho$-tensor correlations, the
balance of attraction and repulsion is changed by the parametrization PKA1, and this constitutes the physical reason
for the improvement of the nuclear shell structure.


\end{document}